\documentclass[a4paper,fleqn]{cas-dc}

\usepackage[numbers]{natbib}

\usepackage{siunitx}
\usepackage{tikz}
\usetikzlibrary{
  arrows,
  arrows.meta,
  automata,
  backgrounds,
  calc,
  decorations,
  decorations.pathmorphing,
  decorations.pathreplacing,
  fit,
  matrix,
  patterns,
  chains,
  shadows,
  shapes,
  shapes.geometric,
  shapes.gates.logic.IEC,
  shapes.gates.logic.US,
}
\usepackage{relsize}

\usepackage{uzlcolor}
\usepackage{fontawesome5}

\small
\newcommand{\score}[2]{
  \cellcolor[HTML]{#1}{#2} %{\textcolor{white}{#2}}
}

\newcommand{\present}{$\bullet$} 
\newcommand{\absent}{--}

\def\tsc#1{\csdef{#1}{\textsc{\lowercase{#1}}\xspace}}
\tsc{WGM}
\tsc{QE}
\begin{document}
\let\WriteBookmarks\relax
\def\floatpagepagefraction{1}
\def\textpagefraction{.001}

\shorttitle{Designing Molecular Communication Networks}

\shortauthors{Waldner et~al.}

\title[mode = title]{Designing DNA-Based Molecular Communication Networks Using Experimentally Validated Components}




\author[uzl-itm]{Andreas Waldner}
\corref{cor1}
\ead{andreas.waldner@uni-luebeck.de}
\cortext[cor1]{Corresponding author}

\author[uzl-itm]{Regine Wendt}
\ead{regine.wendt@uni-luebeck.de}

\author[uzl-itm]{Nils Stumpf}
\ead{n.stumpf@uni-luebeck.de}

\author[uzl-itm]{Sarah Scheer}
\ead{sarah.scheer@uni-luebeck.de}

\author[uzl-itm]{Christian Hyttrek}
\ead{c.hyttrek@uni-luebeck.de}

\author[uzl-itm]{Stefan Fischer}
\ead{stefan.fischer@uni-luebeck.de}

\author[uzl-itm]{Florian-Lennert Lau}
\ead{f.lau@uni-luebeck.de}


\affiliation[uzl-itm]{organization={Universität zu Lübeck},
  addressline={Ratzeburger Allee 160},
  city={Lübeck},
  postcode={23562},
  country={Germany}}

%\cortext[1]{Corresponding author}



\begin{abstract}
  Molecular communication is a novel paradigm for the design of nanonetworks.
  It uses molecules as information carriers and enables communication between nanodevices.
  DNA-based molecular communication is a promising approach for medical applications, such as in-body networks for diagnostics and drug delivery.
  However, the transition from theoretical models to practical experiments is still in its early stages.
  This paper analyzes experimentally validated DNA-tiles, nanocarriers, and biosensors relevant to implementing DNA-tile-based nanonetworks in vitro.
  We evaluate these components based on their stability, compatibility, cost-effectiveness, and overall suitability for wet-lab implementation.
  Based on this analysis, we identify promising component combinations and derive a blueprint for a minimal proof-of-concept experiment.
  The resulting framework provides researchers with both a structured overview of experimentally relevant design choices and a concrete starting point for implementing and validating DNA-based nanonetworks in the wet lab.
\end{abstract}


\begin{highlights}
  \item Analysis of experimentally validated DNA-tiles, nanocarriers, and biosensors for DNA-tile-based nanonetworks.
  \item Evaluation of component complexity, stability, compatibility, and cost-effectiveness.
  \item Derivation of a blueprint for a minimal proof-of-concept experiment.
\end{highlights}

%\nocite{*}

% Keywords
\begin{keywords}
  Molecular communication \sep DNA nanotechnology \sep DNA-based nanonetworks \sep DNA-tiles \sep DNA origami \sep DNA crisscross polymerization \sep nanocarriers \sep biosensors \sep in-body networks
\end{keywords}

\maketitle

\section{Introduction}
\label{sec:intro}

% Abgrenzung zu anderen Surveys/Reviews am besten aus dem Journal

The fields of medicine and computer science have long been interconnected.
Complex and diverse medical systems have been created or improved over decades, ranging from smartwatches tracking sleep, pulse and blood pressure to magnetic resonance imaging for diagnostics.
The persistent miniaturization of computational units over the years led to powerful and miniscule devices that advance not only medicine, but every field of research.
Following this trend of miniaturization, the concept of communicational and computational systems at the nanoscale called \textit{nanonetworks} emerged in 2008 and shaped the nanocommunication sciences since then~\cite{akyildiz2008nanonetworks}.

Nanonetworks consist of nanoscale devices that communicate and cooperate to perform specific tasks.
Nanodevices formally split into nanosensors, nanomachines, nanonodes and nanorobots with different sizes, shapes and functions~\cite{buether2017nanodefinition}.
Akyildiz et al. define three approaches for the development of nanodevices~\cite{akyildiz2008nanonetworks}: Top-down, bottom-up and bio-hybrid.
The top-down approach uses techniques such as electron beam lithography~\cite{tseng2003electronbeamlitho} or micro-contact printing~\cite{lee2004microprinting} to downscale microelectronic devices to nanoscale.
The bottom-up approach builds on electrochemical processes or specific molecules, such as \textit{deoxyribonucleic acid} (DNA), capable of self-assembly~\cite{balzani2006nanomachines}.
The bio-hybrid approach uses existing biological nanomachines as components or models for the development of artificial nanodevices

This work focuses on DNA-based nanonetworks as a bottom-up approach for medical applications, particularly in the context of In-Body Networks~\cite{dressler2015inbody}.
The focus of this work further lies on DNA-based nanonetworks, one bottom-up approach for In-Body networks.
DNA is a molecule consisting of a sugar-phosphate backbone and the four nucleotide bases \textit{adenine} (A), \textit{thymine} (T), \textit{cytosine} (C) and \textit{guanine} (G)~\cite{watson1953dna}.
DNA can exist as \textit{single-stranded DNA} (ssDNA), while complementary strands can form double-stranded DNA through specific base pairing between A and T, and between C and G~\cite{watson1953dna}.
DNA-based nanonetworks use DNA as the main building block for the employed nanodevices.
A more detailed introduction to DNA-based nanonetworks can be found in Section~\ref{sec:prelim}.

DNA-based approaches are particularly promising for medical applications due to the programmability of DNA and its potential for biological compatibility, although stability and immune responses remain important challenges.~\cite{chen2018dna}.
In-Body Networks could revolutionize medical applications, from diagnostics to drug delivery.
Li et al. demonstrate the use of DNA-based nanorobots for targeted drug delivery against specific types of cancer, achieving promising results~\cite{li2018nanorobot}.

Molecular communication has been extensively investigated through various paradigms, including diffusion-based signaling and biologically inspired communication mechanisms~\cite{hiyama2010molecular,malak2012molecular}.
While DNA has been widely studied as a medium for information storage~\cite{cevallos2022storage,nergui2024memory}, its use as a structural carrier of messages that can additionally perform computations represents a comparatively recent research direction.
In particular, DNA-tile-based nanonetworks have been investigated primarily through theoretical models and simulations~\cite{lau2019computation,gomez2024abnormalitydetection,kaussow2026nettas, lau2025offtheshelf, lau2024inflammation}.
Despite these theoretical advances, translating DNA-tile-based nanonetworks into functional experimental systems remains a significant challenge.
While previous research has demonstrated the feasibility of individual DNA nanostructures and molecular mechanisms~\cite{winfree1998tiles}, their integration into complete communication systems requires further wet-lab validation.
This transition from theoretical models to experimental implementation poses considerable interdisciplinary challenges, particularly for computer science researchers.

To address the gap between theoretical models and experimental implementation, this work aims for three main contributions.
First, we identify and structure the main components required for a proof-of-concept experiment and analyze selected implementations of DNA-tiles, nanocarriers, and biosensors.
Second, we develop a systematic scoring methodology to compare these implementations with respect to DNA compatibility, cost efficiency, and experimental feasibility.
Third, based on this evaluation, we derive a concrete experimental blueprint with alternative component candidates and establish a reproducible pipeline for transferring theoretical nanonetwork models toward wet-lab implementation.

Based on that, the following Section~\ref{sec:prelim} defines the basic concepts and terms for our work.
Section~\ref{sec:sota} analyses and scores nanocomponents for their viability in DNA-based nanonetworks.
After that, Section~\ref{sec:blueprint} conceptualizes a blueprint for a minimal proof-of-concept experiment using the most viable nanocomponents from before.
Section~\ref{sec:disc} discusses the concepts and possible challenges.
Lastly, Section~\ref{sec:conclusion} offers a brief overview of the matter and discusses possible future work.

%%%%%%%%%%%%%%%%%%%%%%%%%%%%%%%%%%%%%%%%%%%%%%%%%%%%%%%%%%%%%%%%%%%%%%%%%%%%%%%%%%%%%%%%%%%%%%%
%%%%%%%%%%%%%%%%%%%%%%%%%%%%%%%%%%%%%%%%%%%%%%%%%%%%%%%%%%%%%%%%%%%%%%%%%%%%%%%%%%%%%%%%%%%%%%%
%%% PRELIM %%%%%%%%%%%%%%%%%%%%%%%%%%%%%%%%%%%%%%%%%%%%%%%%%%%%%%%%%%%%%%%%%%%%%%%%%%%%%%%%%%%%
%%%%%%%%%%%%%%%%%%%%%%%%%%%%%%%%%%%%%%%%%%%%%%%%%%%%%%%%%%%%%%%%%%%%%%%%%%%%%%%%%%%%%%%%%%%%%%%
%%%%%%%%%%%%%%%%%%%%%%%%%%%%%%%%%%%%%%%%%%%%%%%%%%%%%%%%%%%%%%%%%%%%%%%%%%%%%%%%%%%%%%%%%%%%%%%

\section{Preliminary Work}
\label{sec:prelim}
\begin{figure}
  \centering
  \includegraphics[width=\linewidth]{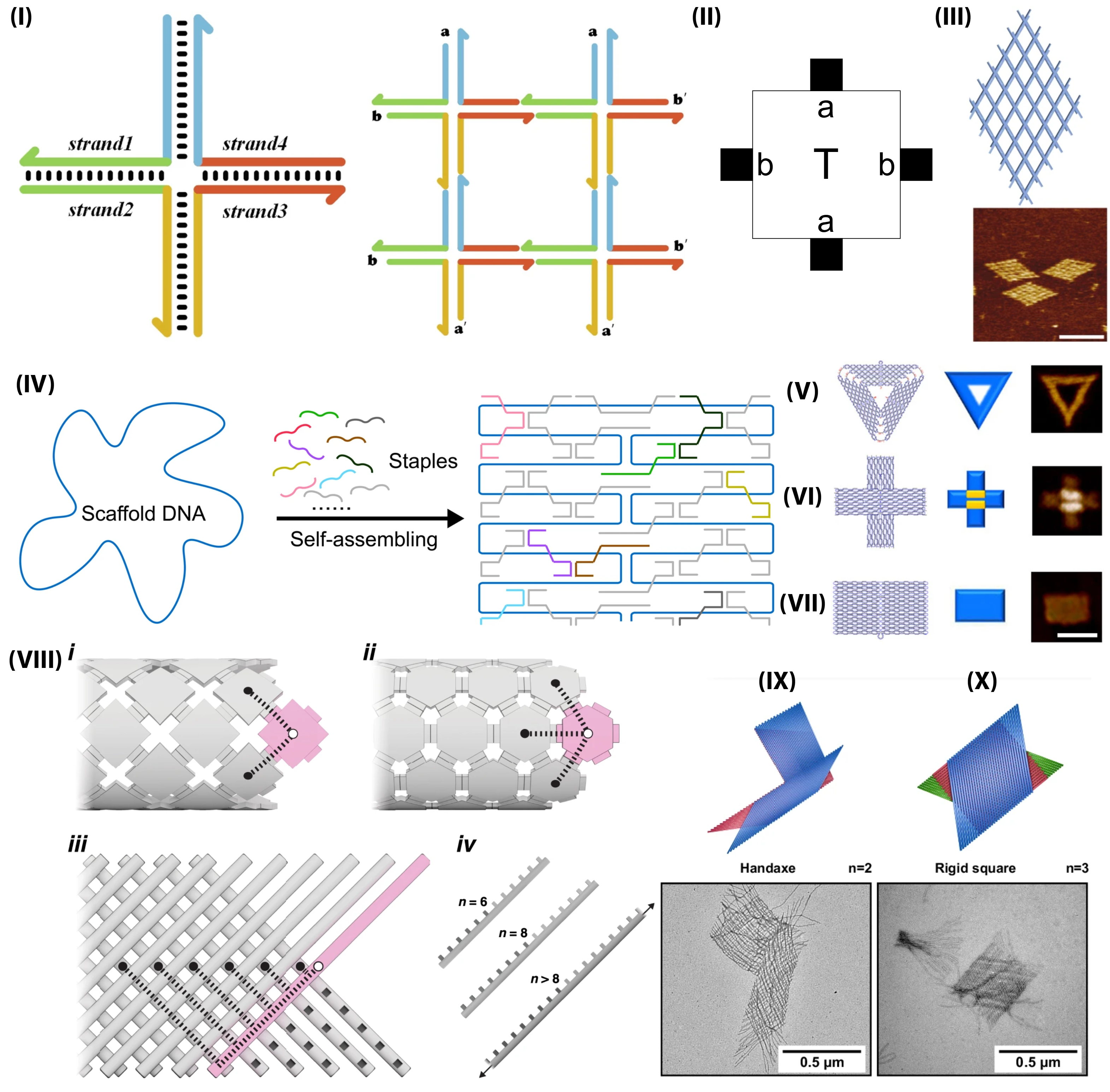}
  \caption{An overview of DNA building blocks. (I) A general schematic of a DNA tile with four arms. Reproduced from~\cite{xu2025dna}, licensed under CC BY 4.0. (II) Mathematical representation of a DNA tile named (T), with glue labels (a) and (b) and glue strengths indicated by black boxes. Based on the tile model in~\cite{lau2019computation}. (III) Model and AFM images of a lattice of DNA tiles. Reproduced from~\cite{wang2019nanostructures}, licensed under CC BY 4.0. (IV) Concept of DNA origami with a long scaffold strand and staples that lock the scaffold into the desired shape. Reproduced from~\cite{zhou2023origami}, licensed under CC BY 4.0. (V--VII) Models and AFM images of triangular, cross-shaped, and block-shaped DNA origami structures. Reproduced from~\cite{zhang2017origami}, licensed under CC BY 4.0. (VIII) Comparison of \texttt{i} square-tile nanotubes, \texttt{ii} hexagonal-tile nanotubes, and \texttt{iii} crisscross polymerization regarding the number of neighbors. In \texttt{iv}, different slat lengths are shown. Reproduced from~\cite{minev2021crisscross}, licensed under CC BY 4.0. (IX--X) Models and TEM images of handaxe and rigid-square structures assembled by crisscross polymerization. Reproduced from~\cite{aquilina2026crisscross}, licensed under CC BY 4.0.}
  \label{fig:DNA_tech}
\end{figure}

This section defines DNA-tiles and DNA-based nanonetworks as the basis of the following sections.
A more detailed description of nanonetworks and specifically DNA-tile-based self-assembly systems can be found in~\cite{lau2024book_nanonetworks, gomez2024abnormalitydetection, lau2022mc, lau2024nanoprotocol, lau2021molcom}.
The following subsection defines the mathematical and natural concept of a DNA-tile.
The second subsection defines DNA-based nanonetworks and the components and terms needed to understand them.

\subsection{DNA-Tile Model and Self-Assembly}
\label{subsec:dna-tiles}
We now define DNA-tiles as nanostructures capable of computation and usable for communication as the main component of DNA-based nanonetworks.
In 1998, Winfree et al. demonstrated DNA-tiles in the form of \textit{double-crossover} (DX) molecules, which self-assemble into two-dimensional crystalline structures through programmable sticky-end interactions~\cite{winfree1998tiles}.
Figure~\ref{fig:DNA_tech} (I) shows a schematic representation of a single tile.
The DNA tile is composed of DNA, but no specific structure or size is required, as a wide range of tile designs can be used.
We use DNA-tile and tile synonymously as other tiles are not regarded in this work.
In many tile-based self-assembly systems, tiles interact through single-stranded DNA overhangs called \textit{sticky ends}, which bind to complementary sequences on other tiles.
The binding between complementary sticky ends depends on their sequences and the environmental conditions, particularly the temperature of the system.
In the mathematical tile model, these dependencies are represented by the sticky ends and their so-called \textit{glue strength}.
The example tile in Figure~\ref{fig:DNA_tech} (I) can connect with itself, because the left to right and the top to bottom base sequences are reverse complementary to each other.
This can be seen in the mathematical representation in Figure~\ref{fig:DNA_tech} (II) as well.
This form helps to understand which sticky ends can bind with which tiles more easily.
The ``T'' in this example is the name of the tile, the ``a'' and ``b'' labels show which sticky ends bind with each other.
The small black boxes in Figure~\ref{fig:DNA_tech} (II) show the glue strength for the sticky end.
In the mathematical model, a tile can attach to an existing assembly if the sum of its glue strength reaches a certain threshold $\tau$ , which is called the \textit{temperature} $\tau$  of the system.
Throughout this work, we follow the notation and terminology introduced by Lau et al.~\cite{lau2019computation} for DNA-tiles, their properties, and their mathematical representation.
A well-designed set of DNA-tiles can be used for more complex structures as a building block or for computational purposes.
Figure~\ref{fig:DNA_tech}~(III) shows an \textit{atomic force microscopy} (AFM) image of a lattice of tiles with four sticky ends that bind with each other.

DNA structures such as DNA-tiles are synthesized and self-assemble into the predefined forms.
Other than basic self-assembly of ssDNA, two methods of building DNA structures have emerged over the years: DNA origami and DNA crisscross polymerization.
DNA origami was introduced by Rothemund in 2006~\cite{rothemund2006origami} and has since become a widely used technique in DNA nanotechnology.
It uses a long ssDNA scaffold and multiple short staple strands to form predefined structures, as shown in Figure~\ref{fig:DNA_tech} (IV).
They are designed and synthesized, such that the staple strands connect and force the scaffold into a specific shape.
Figure~\ref{fig:DNA_tech} (V) to (VII) show exemplary DNA origami shapes and their AFM images.
Crisscross polymerization uses DNA slats that bind to multiple neighboring slats in a crisscross pattern to form a stable structure~\cite{minev2021crisscross}.
These slats can be constructed from single-stranded DNA with multiple short binding domains~\cite{minev2021crisscross} or from DNA origami structures~\cite{wintersinger2023crisscross}.
Figure~\ref{fig:DNA_tech} (VIII) compares the crisscross slats with square- and hexagonal-tile nanotubes~\cite{minev2021crisscross}.
Figure~\ref{fig:DNA_tech} (VIII) i shows that a new square-tile usually binds to two nearest neighbors with a maximum amount of four connections.
Accordingly, the hexagonal-tile in Figure~\ref{fig:DNA_tech} (VIII) ii usually binds to three nearest neighbors with maximal six connections.
In Figure~\ref{fig:DNA_tech} (VIII) iii and iv, the crisscross pattern can be seen.
The number of neighboring components connected to each other can easily be increased from six to eight or more.
Figure~\ref{fig:DNA_tech} (IX) and (X) show exemplary crisscross polymerization models and their \textit{Transmission electron microscopy} (TEM) images.

\subsection{DNA-Based Nanonetworks}
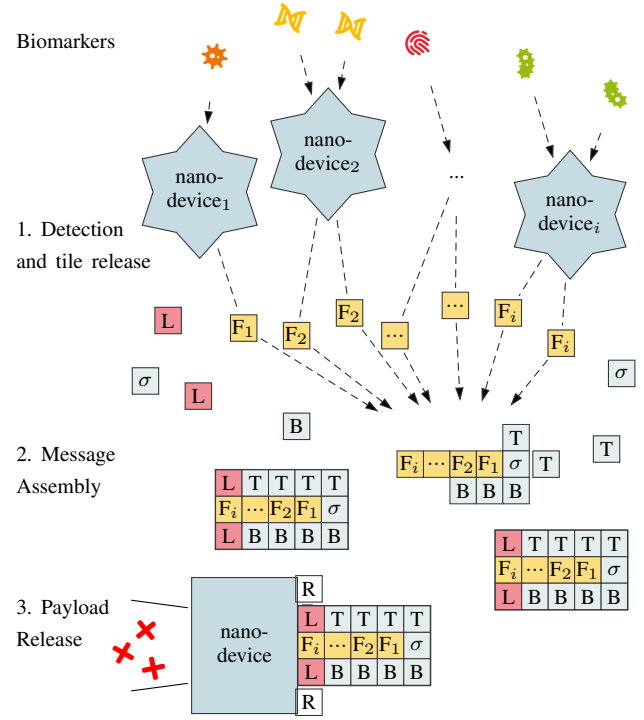
\begin{figure}
  \begin{center}

    \begin{tikzpicture}[scale=1]
      \path [every node/.style={
            anchor=west,
            text width=2.5cm,
          }]

      node at (-0.35,9.7) {\footnotesize Biomarkers}
      node at (-0.35,7) {\footnotesize 1. Detection\\ and tile release}
      %node at (-0.35,7.5) {\footnotesize 2. Tile-\\\footnotesize Release}
      node at (-0.35,4) {\footnotesize 2. Message Assembly\footnotesize}
      node at (-0.35,2) {\footnotesize 3. Payload\\ Release\footnotesize};

      % FPS
      \path[every node/.style={
            shape=star,
            fill=none,
            draw=none,
            inner sep=1mm,
            rotate={rand*360},
          }]
      (2.4, 9.5)  node (marker1) {\textcolor{uzl_orange_2}{\faVirus}}
      (3.4, 10.0) node (marker2) {\textcolor{uzl_yellow_2}{\faDna}}
      (4.2, 9.9)  node (marker3) {\textcolor{uzl_yellow_2}{\faDna}}
      (5.1, 9.66) node (marker4) {\textcolor{uzl_red_2}{\faFingerprint}}
      (6.5, 9.4)  node (marker5) {\textcolor{uzl_green_2}{\faBacterium}}
      (7.7, 9)    node (marker6) {\textcolor{uzl_green_2}{\faBacterium}};

      % nano nodes
      \path[every node/.style={
            shape=star, star points=6, star point height=2.6mm,
            fill=uzl_oceanblue_1,
            draw=black!80,
            font=\smaller,
            inner sep=0mm,
            inner xsep=0mm,
            inner ysep=0mm,
            align=center,
          }]
      (2.2, 7.7) node (sensor1) {nano-\\device$_{1}$}
      (3.9, 8.2) node   (sensor2) {nano-\\device$_{2}$}
      (5.6, 7.9) node[draw = none, fill = white] (sensor3) {...}
      (7.1, 7.4) node (sensor4) {nano-\\device$_{i}$};

      % assembly pieces

      \path[every node/.style={
            draw=black!80,
            font=\smaller,
            anchor=west,
            minimum width=3.5mm,
            minimum height=3.5mm,
            inner sep=0mm,
          }]
      (2.6, 5.9) node [fill=uzl_yellow_2!50] (marker1a) {F$_1$}
      (3.3, 5.8) node [fill=uzl_yellow_2!50] (marker2a) {F$_2$}
      (4,   6.1) node [fill=uzl_yellow_2!50] (marker2b) {F$_2$}
      (4.6, 5.8) node [fill=uzl_yellow_2!50] (marker3a) {$...$}
      (5.4, 6.2) node [fill=uzl_yellow_2!50] (marker3b) {$...$}
      (6.1, 6.1) node [fill=uzl_yellow_2!50] (marker4a) {F$_i$}
      (6.8, 5.7) node [fill=uzl_yellow_2!50] (marker4b) {F$_i$}
      (1.3, 5.2) node [fill=uzl_oceangreen_10] (free1) {$\sigma$}
      (1.6, 6)   node [fill=uzl_red_2!50] (free6) {L}
      (2,   5)   node [fill=uzl_red_2!50] (free2) {L}
      (6.6, 4.1) node [fill=uzl_oceangreen_10] (free5) {T}
      (7.4, 4.3) node [fill=uzl_oceangreen_10] (free3) {T}
      (7.6, 5.3) node [fill=uzl_oceangreen_10] (free4) {$\sigma$}
      (3.3, 4.6) node [fill=uzl_oceangreen_10] (free7) {B};

      % assembled messages

      \begin{scope}[nodes={
              draw=black!80,
              line width=0.1mm,
              font=\smaller,
              anchor=west,
              minimum width=3.5mm,
              minimum height=3.5mm,
              inner sep=0mm,
              column sep=-0.1mm,
              row sep=-0.1mm,
            }]

        \foreach \x/\y/\n in {2.4/3.5/1, 6.1/2.7/2} {
            \path
            (\x, \y) node (msg\n) [matrix of nodes, ampersand replacement=\&,
              row 1 column 1/.style={nodes={fill=uzl_red_2!50}},
              row 1 column 2/.style={nodes={fill=uzl_oceangreen_10}},
              row 1 column 3/.style={nodes={fill=uzl_oceangreen_10}},
              row 1 column 4/.style={nodes={fill=uzl_oceangreen_10}},
              row 1 column 5/.style={nodes={fill=uzl_oceangreen_10}},
              row 2 column 1/.style={nodes={fill=uzl_yellow_2!50}},
              row 2 column 2/.style={nodes={fill=uzl_yellow_2!50}},
              row 2 column 3/.style={nodes={fill=uzl_yellow_2!50}},
              row 2 column 4/.style={nodes={fill=uzl_yellow_2!50}},
              row 2 column 5/.style={nodes={fill=uzl_oceangreen_10}},
              row 3 column 1/.style={nodes={fill=uzl_red_2!50}},
              row 3 column 2/.style={nodes={fill=uzl_oceangreen_10}},
              row 3 column 3/.style={nodes={fill=uzl_oceangreen_10}},
              row 3 column 4/.style={nodes={fill=uzl_oceangreen_10}},
              row 3 column 5/.style={nodes={fill=uzl_oceangreen_10}},] {
              L \& T \& T \& T \& T \\
              F$_i$ \& $...$ \& F$_2$ \& F$_1$ \& $\sigma$ \\
              L \& B \& B \& B \& B \\
            };

            % % binding receptangles.
            % \path
            %     [fill=black!80]
            %     (msg\n) ++(-5mm,0)
            %     ++(0, 3.75mm)
            %     ++(-0.8mm, 0) rectangle +(1.4mm, 0.7mm)
            %     ++(0, -7.5mm) rectangle +(1.4mm, -0.7mm);
          }

        % partially assembled message

        \path (4.8, 4.1) node (msg3) [matrix of nodes,draw=none,
          row 1 column 1/.style={nodes={fill=uzl_red_2!50}},
          row 1 column 2/.style={nodes={fill=uzl_oceangreen_10}},
          row 1 column 3/.style={nodes={fill=uzl_oceangreen_10}},
          row 1 column 4/.style={nodes={fill=uzl_oceangreen_10}},
          row 1 column 5/.style={nodes={fill=uzl_oceangreen_10}},
          row 2 column 1/.style={nodes={fill=uzl_yellow_2!50}},
          row 2 column 2/.style={nodes={fill=uzl_yellow_2!50}},
          row 2 column 3/.style={nodes={fill=uzl_yellow_2!50}},
          row 2 column 4/.style={nodes={fill=uzl_yellow_2!50}},
          row 2 column 5/.style={nodes={fill=uzl_oceangreen_10}},
          row 3 column 1/.style={nodes={fill=uzl_red_2!50}},
          row 3 column 2/.style={nodes={fill=uzl_oceangreen_10}},
          row 3 column 3/.style={nodes={fill=uzl_oceangreen_10}},
          row 3 column 4/.style={nodes={fill=uzl_oceangreen_10}},
          row 3 column 5/.style={nodes={fill=uzl_oceangreen_10}},] {
          &   &   &   & T \\
          F$_i$ & $...$ & F$_2$ & F$_1$ & $\sigma$ \\
          &   & B & B & B \\
        };

      \end{scope}

      % Nanomachines

      \foreach \x/\y/\n in {3.5/1.7/1} {

          \path [every node/.style={
                fill=uzl_oceanblue_1,
                draw=black!80,
                font=\smaller,
                align=center,
                anchor=east,
                minimum height=18mm,
                minimum width=14mm,
                inner sep=0mm,
              }]
          (\x, \y) node (machine\n) {nano-\\device};

          % receptor Xs
          \path [every node/.style={
                draw=black!80,
                font=\smaller,
                align=center,
                minimum width=3.5mm,
                minimum height=3.5mm,
                inner sep=0mm,
              }]
          (\x, \y) ++(1.45mm, 7.5mm)  node (mRecU\n) {R}
          (\x, \y) ++(1.45mm, -7.5mm) node (mRecL\n) {R};

          %% binding receptangles on receptor Xs
          \path [fill=black!80, anchor=west]
          (\x, \y) ++(1.25mm, 4.75mm) ++(-0.7mm, 0) rectangle +(1.4mm, 0.7mm)
          (\x, \y) ++(1.25mm, -4.55mm) ++(-0.7mm, 0) rectangle +(1.4mm, -0.8mm);
        }

      % Machine 2 molecule release.

      \path [draw]
      (machine1) ++(-0.75,0.5) -- +(-0.75, 0.1)
      ++(-0,   -1)  -- +(-0.75, -0.1);

      \path [every node/.style={
            shape=isosceles triangle,
            minimum width=0.0mm,
            isosceles triangle apex angle=60,
            inner sep=0.6mm,
            fill=none,
            draw=none,
            rotate={rand*360},
          }]
      (machine1) ++(-0.5,0)
      node at +(-0.8, 0.2) (rel1) {\textcolor{red}{\faPlus}}
      node at +(-1.1, -0.1) (rel2) {\textcolor{red}{\faPlus}}
      node at +(-0.7, -0.3) (rel3) {\textcolor{red}{\faPlus}};

      % Bound Message in Machines
      \foreach \pos/\n in {machine1/5,machine1/5} {
          \path [nodes={
                draw=black!80,
                line width=0.1mm,
                font=\smaller,
                anchor=west,
                minimum width=3.5mm,
                minimum height=3.5mm,
                inner sep=0mm,
                column sep=-0.1mm,
                row sep=-0.1mm,
              }]
          {(\pos.east) + (-0.01,0)} node (msg\n)
          [matrix of nodes, ampersand replacement=\&,
            row 1 column 1/.style={nodes={fill=uzl_red_2!50}},
            row 1 column 2/.style={nodes={fill=uzl_oceangreen_10}},
            row 1 column 3/.style={nodes={fill=uzl_oceangreen_10}},
            row 1 column 4/.style={nodes={fill=uzl_oceangreen_10}},
            row 1 column 5/.style={nodes={fill=uzl_oceangreen_10}},
            row 2 column 1/.style={nodes={fill=uzl_yellow_2!50}},
            row 2 column 2/.style={nodes={fill=uzl_yellow_2!50}},
            row 2 column 3/.style={nodes={fill=uzl_yellow_2!50}},
            row 2 column 4/.style={nodes={fill=uzl_yellow_2!50}},
            row 2 column 5/.style={nodes={fill=uzl_oceangreen_10}},
            row 3 column 1/.style={nodes={fill=uzl_red_2!50}},
            row 3 column 2/.style={nodes={fill=uzl_oceangreen_10}},
            row 3 column 3/.style={nodes={fill=uzl_oceangreen_10}},
            row 3 column 4/.style={nodes={fill=uzl_oceangreen_10}},
            row 3 column 5/.style={nodes={fill=uzl_oceangreen_10}},] {
            L \& T \& T \& T \& T \\
            F$_i$ \& $...$ \& F$_2$ \& F$_1$ \& $\sigma$ \\
            L \& B \& B \& B \& B \\
          };
        }

      % movement arrows

      \path [draw=black!80,
      line width=0.15mm, densely dashed,
      >={Triangle[length=1.8mm, width=1.1mm]}, ->,
      shorten <=0.5mm,
      shorten >=1mm,
      ]
      (marker1) edge (sensor1)
      (marker2) edge (sensor2)
      (marker3) edge (sensor2)
      (marker4) edge (sensor3)
      (marker5) edge (sensor4)
      (marker6) edge (sensor4);

      \path [draw=black!80,
        line width=0.15mm, densely dashed,
        shorten <=1mm,
        shorten >=0.5mm,
      ]
      (sensor1) edge (marker1a)
      (sensor2) edge (marker2a)
      (sensor2) edge (marker2b)
      (sensor3) edge (marker3a)
      (sensor3) edge (marker3b)
      (sensor4) edge (marker4a)
      (sensor4) edge (marker4b);

      \path [draw=black!80,
      line width=0.15mm, densely dashed,
      >={Triangle[length=1.8mm, width=1.1mm]}, ->,
      shorten <=0.5mm,
      shorten >=3mm,
      ]
      (marker1a) edge (msg3)
      (marker2a) edge (msg3)
      (marker2b) edge (msg3)
      (marker3a) edge (msg3)
      (marker3b) edge (msg3)
      (marker4a) edge (msg3)
      (marker4b) edge (msg3);

    \end{tikzpicture}
    \caption{Example of a DNA-tile-based nanonetwork. A not further specified group of nanodevices detects multiple kinds of biomarkers. When detecting their designated biomarkers, the nanodevices release tiles into the system. Together with tiles that already exist in the medium, the tiles self-assemble into a message molecule. Only when the whole message molecule can assemble, the ligands and receptors bind correctly and an additional nanodevice detects the message molecule. This nanodevice releases its payload, which can contain another set of tiles, medical drugs or other nanocomponents. Design derived from~\cite{lau2019computation}.}
    \label{fig:nanonetwork}
  \end{center}
\end{figure}

This subsection defines DNA-based nanonetworks and important components, processes and constraints related to them.
Based on the previously introduced building blocks, we describe the operating principle of a DNA-based nanonetwork following Lau et al.~\cite{lau2019computation}.
Figure~\ref{fig:nanonetwork} illustrates the corresponding sequence, which consists of the following steps:
In the first step of Figure~\ref{fig:nanonetwork}, we assume, that there are multiple different nanodevices.
These nanodevices are able to detect different types of biomarkers and distinguish between biomarkers with minor differences.
We further assume, that these nanodevices hold a certain number of tiles, which are released when the specific biomarker is detected.
Starting from a seed-tile~$\sigma$ a \textit{message molecule} self-assembles, in the second step of Figure~\ref{fig:nanonetwork}.
The tiles are designed, such that the red ``L'' tiles only bind, when the whole message molecule assembles beforehand.
The ``L'' tiles serve as ligands that can bind to the corresponding receptor element ``R'' of the nanodevice in the third step of Figure~\ref{fig:nanonetwork}.
This final nanodevice opens up and releases its payload, only if a complete message molecule binds to both of its receptors.
The payload of the nanodevice can either be more nanodevices, tiles, or it could be used for drug release in the medical context.

Both types of nanodevices in Figure~\ref{fig:nanonetwork} need some kind of recognition agent, more commonly called \textit{biosensor} in the medical context.
Various molecules can serve as recognition elements in biosensors if they interact selectively with a specific target molecule.
To function as a biosensor, this interaction must produce a detectable signal.
Naturally, at the nanoscale there are biosensors, whose reaction is so miniscule that it can be considered immeasurable.
These types of biosensors use additional signal amplification to enhance their detectability~\cite{loretan2020origamisensor,abolhasan2019hairpin}.

The specific design of the employed nanodevices in a DNA-based nanonetwork depends on their use case and context.
A general approach is to use nanocarriers with locking and opening mechanisms~\cite{andersen2009boxlid}.
\textit{Nanocarriers} are nanoscale systems designed to transport and protect cargo, such as drugs or other molecules.
Depending on their material and design, cargo can be encapsulated within the carrier or bound to its structure.
They are three-dimensional constructs with sizes ranging from \SIrange{1}{1000}{\nm}~\cite{gressler2025nanocarrier}.
Nanocarriers do not need to be biocompatible by default, but for the medical context and this work, it is an important selection criterion.
One widely used method of constructing complex DNA structures is called thermal annealing~\cite{rothemund2006origami}.
In this work, we use the term \textit{annealing} synonymously with thermal annealing.
In the annealing process the ssDNA components are mixed and heated up over a short period of time.
After that, it is cooled down over a long period of time.
In the process of heating up, existing base pairings are disrupted and secondary structures and unwanted hybridizations melt.
While cooling down slowly, the strands self-assemble into the intended structures.
Castro et al. describe their annealing processes in high details for their DNA origami construction~\cite{castro2011primer}.
They briefly heated up the mixtures to \SI{80}{\degreeCelsius}.
The cooling process to \SI{20}{\degreeCelsius} is prolonged from two hours to seven days, depending on the complexity of the structure and the amount of DNA origami layers.
Yan et al. describe the annealing of DNA-tiles by cooling the mixture from \SI{90}{\degreeCelsius} to room temperature over four hours or longer~\cite{yan2003tileanneal}.

DNA in single- and multi-stranded forms has different levels of stability and degradation based on human body conditions such as the temperature.
The melting temperature in Celsius ($T_m$) of short DNA strands can be estimated using the Wallace rule~\cite{wallace1979oligonucleotide}:
\begin{equation}
  T_m = 2 (\text{A} + \text{T}) + 4 (\text{G} + \text{C})
\end{equation}
Where A, T, G, and C are the counts of the respective bases in the DNA strand.
This approximation is only a rough estimate and does not consider the influence of the surrounding environment, such as salt concentration or pH value.
For In-body applications, DNA structures must remain stable under the expected conditions.
The sequence design and temperature constraints need to consider the stability of the DNA-tile and the binding properties of the sticky ends simultaneously.
More complex DNA structures with higher temperature constraints can be pre-assembled outside the medium and then introduced into the system.
The design process needs to ensure both stability and functionality.

%%%%%%%%%%%%%%%%%%%%%%%%%%%%%%%%%%%%%%%%%%%%%%%%%%%%%%%%%%%%%%%%%%%%%%%%%%%%%%%%%%%%%%%%%%%%%%%
%%%%%%%%%%%%%%%%%%%%%%%%%%%%%%%%%%%%%%%%%%%%%%%%%%%%%%%%%%%%%%%%%%%%%%%%%%%%%%%%%%%%%%%%%%%%%%%
%%% SOTA %%%%%%%%%%%%%%%%%%%%%%%%%%%%%%%%%%%%%%%%%%%%%%%%%%%%%%%%%%%%%%%%%%%%%%%%%%%%%%%%%%%%%%
%%%%%%%%%%%%%%%%%%%%%%%%%%%%%%%%%%%%%%%%%%%%%%%%%%%%%%%%%%%%%%%%%%%%%%%%%%%%%%%%%%%%%%%%%%%%%%%
%%%%%%%%%%%%%%%%%%%%%%%%%%%%%%%%%%%%%%%%%%%%%%%%%%%%%%%%%%%%%%%%%%%%%%%%%%%%%%%%%%%%%%%%%%%%%%%

\section{Analysis}
\label{sec:sota}

This section examines the state of the art of the three key components required to experimentally realize the DNA-tile-based nanonetwork introduced in Figure~\ref{fig:nanonetwork}: DNA-tiles, nanocarriers, and biosensors.
By analyzing these components individually, we identify suitable building blocks for translating the previously introduced conceptual nanonetwork into a minimal experiment.
For each component, a dedicated subsection analyzes selected approaches and evaluates their suitability for implementing DNA-tile-based nanonetworks.
The analysis primarily focuses on compatibility with in vitro experiments for a minimal \textit{proof of concept} (PoC).
Since future in vivo applications remain a long-term objective, we focus on components with documented use in biological environments or evidence suggesting potential biocompatibility.
However, their suitability for specific in vivo applications requires further experimental validation.
The analysis is limited to selected component implementations relevant to the proposed experimental setup and does not aim to provide an exhaustive review of all available approaches.

We select five evaluation criteria and give a score between one and five for each component.
The most fitting implementation is determined according to the following five parameters, that mainly concern in vitro experiments:

\begin{itemize}
  \item \textit{Amount of Subcomponents} (AoS) needed for a single implementation of the given component
  \item \textit{Amount of Operations} (AoO) needed for the component to be built in a wet-lab environment
  \item \textit{Compatibility with DNA} (CwD) of the given component.
  \item \textit{Level of Stability} (LoS) of the given component in a wet-lab environment.
  \item \textit{Cost of Component} (CoC) of the given component.
\end{itemize}

These criteria were selected to address the practical challenges of implementing a minimal wet-lab experiment, including component complexity, experimental effort, DNA compatibility, stability, and cost
They address the factors that determine the efficiency, feasibility and practical applicability of in vitro experiments.

The AoS shows the number of subcomponents needed for the component to work.
More subcomponents mean more possible points of failure, which could make an otherwise working experiment fail.
We define the AoS as the quantity of different molecules.
Repeated instances of the same building block are counted only once.
If a nanoparticle needs many slightly different molecules, for example different DNA strands, we count every molecular species as one subcomponent.
The scores for AoS are:
\begin{description}
  \item[5] for 1--2 subcomponents,
  \item[4] for 3--5 subcomponents,
  \item[3] for 6--10 subcomponents,
  \item[2] for 11--50 subcomponents and
  \item[1] for 51 or more subcomponents.
\end{description}

We assigned a score of 5 to systems with 1--2 subcomponents which is reflecting a modular and minimal design with less complexity.
A score of 4 applies to 3--5 subcomponents, indicating a level with manageable complexity for the minimal PoC.
For 6--10 subcomponents the system remains feasible but shows increased risk of failure.
The last two scores are given to highly complex systems and therefore huge challenges.
Systems with more than 50 distinct subcomponents receive the lowest score because their preparation and handling are less suitable for the minimal experimental setup considered in this work.

Another key consideration is minimizing the number of laboratory operations to reduce errors.
The AoO expresses the number of steps needed for the subcomponents to be built into the final component.
Operations in this context include, for example, annealing processes, the construction of subcomponents, and other experimental steps required to obtain the final structure.
Each execution of a required laboratory procedure is counted as one operation, even if the same procedure is repeated multiple times.
Each additional step can introduce further sources of error, so the number of operations should be kept as low as possible.
Together, AoS and AoO model the complexity of the nanostructure.
The AoO score is divided into:
\begin{description}
  \item[5] for 0 operations,
  \item[4] for 1--2 operations,
  \item[3] for 3--5 operations,
  \item[2] for 6--9 operations and
  \item[1] for 10 or more operations.
\end{description}

We assigned a score of 5 to systems with 0 operations which is reflecting a ready to use component with no extra effort.
A score of 4 applies to 1--2 operations, reflecting a standard workflow with minimal effort.
For 3--5 operations the system remains feasible for most lab setups.
The last two scores are given to very complex systems and therefore requiring a lot of extra steps which are too complex for the standard lab.

The CwD expresses the level of ease for the component to be used in a DNA dominated domain.
Since DNA is the primary building material of the proposed nanonetwork, components that can directly interact with DNA require less additional functionalization and reduce the complexity of integrating different nanocomponents.
The CwD score is defined as:
\begin{description}
  \item[5] for DNA-native components,
  \item[4] for compatibility with minimal adaptation or addition of single strands to combine DNA with the component,
  \item[3] for compatibility via defined molecules or structures,
  \item[2] for single points of compatibility and
  \item[1] for incompatibility.
\end{description}

A score of 5 is assigned to DNA-native components, as they can be integrated through programmable DNA interactions without requiring additional material interfaces.
A score of 4 applies to minimal adaptations, indicating a level of manageable effort.
For a score of 3, we define compatibility through specific molecules or structures that require additional engineering.
A score of 2 stands for single points of compatibility, limiting reusability and flexibility.
The last score is given to incompatible systems.

The LoS describes the stability of components.
We quantify stability by the half-life $t_{50}$ of the functional component and the melting temperature $T_m$.
If $t_{50}$ or $T_m$ is not reported, we only derive conservative bounds from the experimental details.
We assign the lower score, if both values are reported and differ in their score.
If neither value is reported or can be derived, we score the component based on other stability indicators reported in the specific instance and mark it as an estimate with low evidence.
The scores for LoS are:
\begin{description}
  \item[5] $t_{50}\geq 7$ days or $T_m\geq\SI{75}{\degreeCelsius}$,
  \item[4] $t_{50}\geq\SI{24}{\hour}$ or $T_m\geq\SI{60}{\degreeCelsius}$,
  \item[3] $t_{50}\geq\SI{6}{\hour}$ or $T_m\geq\SI{45}{\degreeCelsius}$,
  \item[2] $t_{50}\geq \SI{1}{\hour}$ or $T_m\geq\SI{30}{\degreeCelsius}$,
  \item[1] $t_{50}<\SI{1}{\hour}$ or $T_m < \SI{30}{\degreeCelsius}$.
\end{description}

We assigned a score of 5 to systems that last several days or have a high melting temperature of $T_m\geq\SI{75}{\degreeCelsius}$.
A score of 4 applies to 24 hours or $T_m\geq\SI{60}{\degreeCelsius}$, indicating a level with manageable complexity for the minimal PoC.
For 6 hours or $T_m\geq\SI{45}{\degreeCelsius}$ the system remains feasible but shows increased risk of failure due to time restrictions.
The last two scores are given to very short living systems or are only viable for immediate use and fast transport.
We defined the thresholds as practical reference values for comparing components within the scope of this paper.
They are not intended as universal measures of biological stability.

Finally, the CoC considers the financial costs of the component.
For a minimal proof-of-concept experiment low costs are important, to maximize the number of experiments given a limited budget.
Since the analyzed component classes differ substantially and are not necessarily offered by the same suppliers, a comprehensive market comparison is outside the scope of this work.
For each component, we therefore use one commercially available product or supplier that provides a sufficiently representative implementation of the required component.
Suppliers were not selected to minimize the price, and the reported values should consequently not be interpreted as market minima or average market prices.
Instead, they serve as reference prices to estimate the approximate cost category of a component.

The CoC score is divided into:
\begin{description}
  \item[5] for components that cost less than €50,
  \item[4] for components that cost from €50 to less than €100,
  \item[3] for components that cost from €100 to less than €200,
  \item[2] for components that cost from €200 to less than €500, and
  \item[1] for components that cost €500 or more.
\end{description}

We assigned a score of 5 to components that cost less than €50, which is high cost-efficiency and availability.
A score of 4 applies to components that cost between €50 and less than €100, indicating a level of manageable cost.
For from €100 to less than €200 the system remains feasible but shows increased costs for most academic labs.
The last two scores are given to high-cost components which therefore mean challenges for standard laboratory experiments and often require more funding.

The following Subsection~\ref{subsec:DNA_tiles} discusses DNA-tiles as the main component of the DNA-based nanonetwork concept first.
Secondly in Subsection~\ref{subsec:nanocarriers}, we view different nanocarrier approaches.
Lastly in Subsection~\ref{subsec:biosensors}, we introduce implementations for biosensors in the medical field.
Each subsection first describes the considered component types individually, with each paragraph focusing on one specific implementation and its relevant properties.
After this overview, the same component types are evaluated individually according to the defined criteria, and the assigned scores are explained.
A corresponding table summarizes the selected component types and their scores and serves as a reference throughout the analysis.
The overall score for each component type is calculated as the sum of the five individual scores, giving each criterion equal weight.

%%%%%%%%%%%%%%%%%%%%%%%%%%%%%%%%%%%%%%%%%%%%%%%%%%%%%%%%%%%%%%%%%%%%%%%%%%%%%%%%%%%%%%%%%%%%%%%
%%% SOTA Tiles %%%%%%%%%%%%%%%%%%%%%%%%%%%%%%%%%%%%%%%%%%%%%%%%%%%%%%%%%%%%%%%%%%%%%%%%%%%%%%%%
%%%%%%%%%%%%%%%%%%%%%%%%%%%%%%%%%%%%%%%%%%%%%%%%%%%%%%%%%%%%%%%%%%%%%%%%%%%%%%%%%%%%%%%%%%%%%%%

\subsection{DNA-Tiles}
\label{subsec:DNA_tiles}

\begin{figure*}
  \centering
  \includegraphics[width=\textwidth]{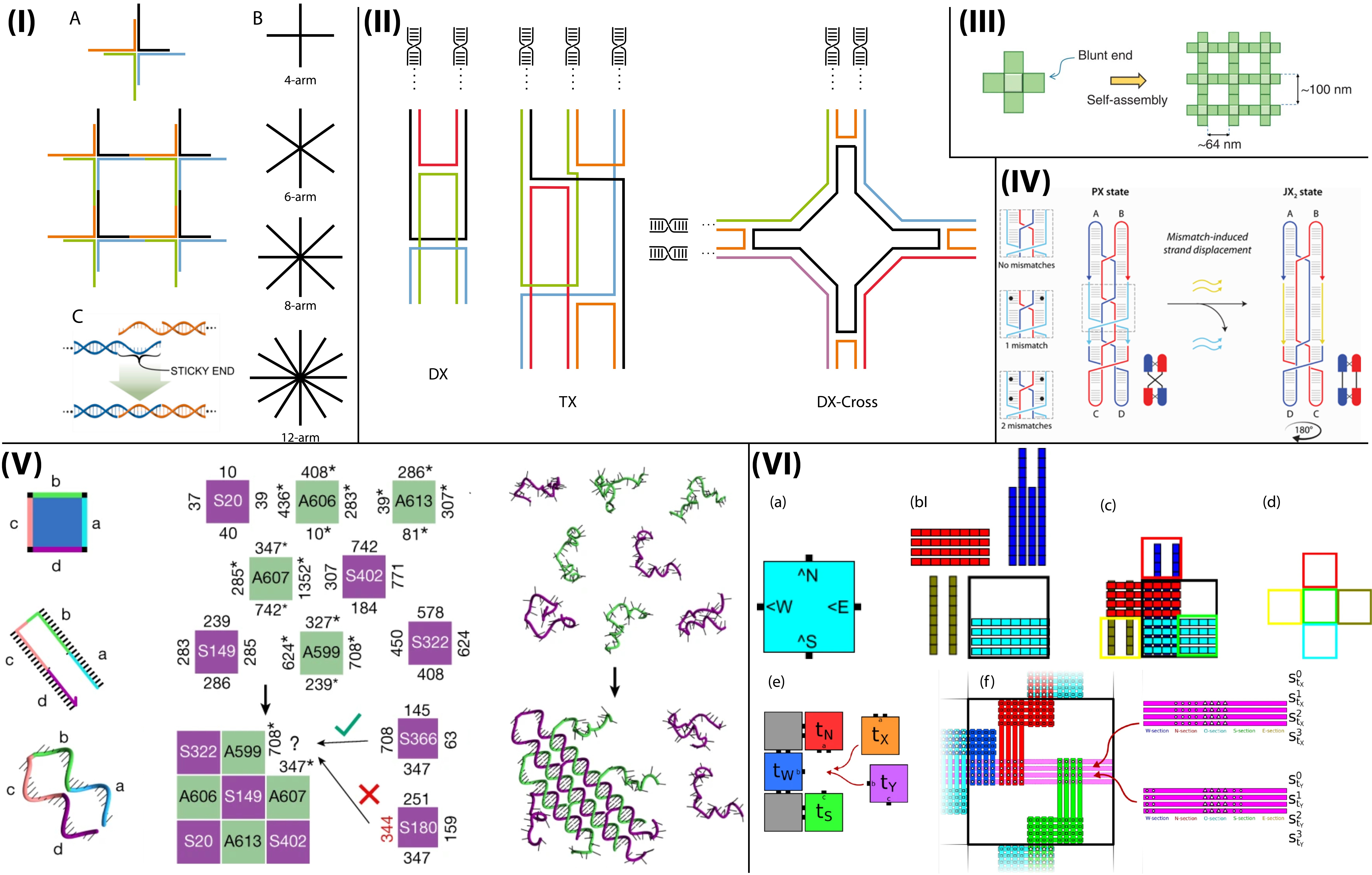}
  \caption{Overview and examples of different kinds of DNA nanostructures suitable for tile implementation. (I) Concept of basic DNA junctions. (A) A simple four-armed junction and a lattice formed by such junctions. (B) Concept of multi-armed junctions analogous to the four-armed junction with the illustration of four-armed, six-armed, 8-armed and twelve-armed junctions. (C) Sticky-end depiction reproduced from~\cite{parikka2021lattice}, licensed under CC BY 4.0. (II) Concept of DX, TX, and DX-cross tiles based on the designs described in~\cite{labean2000txtiles}. (III) Examples of DNA origami tiles, reproduced from~\cite{suzuki2015origamitile}, licensed under CC BY 4.0. (IV) Concept of PX and JX$_2$ tiles and their interconversion, reproduced from~\cite{talbot2025px}, licensed under CC BY 4.0. (V) Exemplary ssDNA-based tile structures, reproduced from~\cite{evans2024ssts}, licensed under CC BY 4.0. On the left, the four domains of the DNA strands are shown and mapped onto a square tile representation. The middle illustrates a simplified mathematical model of the self-assembly process of these tiles. On the right, a possible DNA structure illustrates what such an assembly could look like at the molecular level. (VI) Concept of simulating DNA-tile assembly using crisscross slats. (a) Mathematical representation of a single tile, (b) the corresponding slats separated into functional groups, (c) the assembled slat-based tile, and (d) the representation of its functional regions and output glues. (e) Mathematical representation of a tile assembly with two candidates that can attach through different glue configurations, and (f) the corresponding slat-based simulation with competing decision slats. Panels (a--f) are adapted from~\cite{drake2026slats}, licensed under CC BY 4.0.}
  \label{fig:tiles}
\end{figure*}

As one of the three core components of DNA-based nanonetworks, DNA-tiles are of particular importance because they assemble into the message molecules, which are the basis of the proposed system.
These message molecules not only carry information but can also perform computational operations.
Since different DNA-tile designs vary in properties, this subsection examines potential candidates and assesses their suitability for the proof-of-concept implementation.
Multiple tiles can self-assemble with each other via their sticky ends, when the temperature constraints are met.
Multiple connected tiles are called an \textit{assembly}.
There are many reviews, such as~\cite{heuer-jungemann2019dna, song2017nanostructures, parikka2021lattice, zhan2023dnaorigami, fu2021nanostructures, liu2021concepts}, that show a big variety of DNA-tiles.
A comprehensive tutorial on the design of DNA nanostructures and their sticky ends was published by Piantanida et al.~\cite{piantanida2024tutorial}.
Chen et al.~\cite{chen2020scaling} review DNA-tiles for their scalability in bigger structures.
The following paragraphs first introduce the considered DNA-tile implementations and their relevant properties before assessing their suitability according to the defined metrics.

Figure~\ref{fig:tiles} shows a collection of different DNA-tile designs.
Figure~\ref{fig:tiles}~(I) depicts one of the first tile designs: the Holliday junction~\cite{kallenbach1983junction}.
The four-arm junction in (A) was introduced by Seeman in 1982~\cite{seeman1982junctions}.
It consists of four ssDNA strands, and Figure~\ref{fig:tiles}~(I) illustrates its structure.
DNA-tiles based on Holliday junctions are easy to design and just as easily adaptable.
The six-arm design in Figure~\ref{fig:tiles}~(I)~(B) and a five-arm design were introduced by~\cite{wang1991fivearm}.
The eight- and twelve-arm design were introduced by~\cite{wang2007eightarm}.
All those junctions have the tile specific sticky ends, as seen in Figure~\ref{fig:tiles}~(I)~(A) at the bottom.

More complex and more stable tile implementations are multi-crossover-tiles~\cite{lin2006complex}.
Fu and Seeman introduced DNA \textit{double-crossover} (DX) molecules in 1993~\cite{fu1993dx}.
Winfree et al. later demonstrated their programmable self-assembly into two-dimensional lattices~\cite{winfree1998tiles}.
DX-tile molecules are described with three to four letters~\cite{sa-ardyen2003dx}.
The first letter is always ``D'' for double crossover.
The second letter stands for the relative orientation of their two double-helical domains, with ``A'' for antiparallel and ``P'' for parallel.
The third letter describes the number of helical half-turns between the two crossovers, with ``O'' for odd numbered and ``E'' for even numbered half-turns.
The fourth letter is only needed for DPO molecules, as the extra half-turn can lead to an wide groove separation ``W'' or a narrow groove separation ``N''.
The five different DX-molecules therefore are: DAO, DAE, DPE, DPOW and DPON~\cite{fu1993dx}.
A schematic for a DX-tile can be seen on the left in Figure~\ref{fig:tiles}~(II).
\textit{Triple crossover} (TX)-tiles were introduced by LaBean et al.~\cite{labean2000txtiles}.
Their usual molecules are TAO and TAE, after the same nomenclature from before, but LaBean et al. speak of parallel TX-tiles as well~\cite{labean2000txtiles}.
Figure~\ref{fig:tiles}~(II) shows a TX-tile schematic in the middle.
The DX-Cross tile on the right side of Figure~\ref{fig:tiles}~(II) illustrates how the multi-crossover tile design principle can be extended further.
TX-tiles and multi-crossover tiles with more helices can also be bundled.
The concept of multi-crossover tiles can be extended to multiple different shapes.
Figure~\ref{fig:tiles} (IV) shows another pair of multi-crossover tiles with the \textit{paranemic crossover} (PX)-tile and its topoisomere JX$_2$-tile, designed by Yan et al.~\cite{yan2002px}.
The JX$_2$-tile can be viewed as the PX-tile, with two crossover sections removed.
Accordingly, a JX$_1$-tile is another topoisomere of the PX-tile, with one crossover section removed.
There are multiple reviews that discuss different triplex or even quadruplex DNA structures, such as~\cite{chandrasekaran2018triplex, tang2025triplex, yang2020g-quadruplex}.

Beyond junction- and crossover-based motifs, DNA origami provides another design principle for constructing tile-like DNA nanostructures~\cite{liu2011crystalline}.
As shown in Figure~\ref{fig:tiles}~(III) tiles build with DNA origami are analogously designed to other tile modules.
Their sticky ends are implemented through appropriately designed extensions of the scaffold or staple strands.

In contrast to these multi-stranded tile architectures, \textit{single-stranded tiles} (SSTs) follow a more minimal design principle in which individual DNA strands act as the basic tile units~\cite{yin2008origami_tile}.
Figure~\ref{fig:tiles}~(V) illustrates the concept of SSTs.
A single strand of DNA defined as a SST is typically separated into domains, that bind to domains of other SSTs respectively.
In Figure~\ref{fig:tiles}~(V) on the left side each SST is divided into four domains that bind to up to four different SSTs and therefore assemble into a stable predefined structure.
They can be easily abstracted to a mathematical model similar to Figure~\ref{fig:DNA_tech}~(II).
A very closely related design principle for DNA nanostructures is the DNA brick introduced by~\cite{ke2012dnabricks}.
DNA bricks are single-stranded DNA building blocks used to form three-dimensional structures.
Their modular design is analogous to LEGO\textsuperscript{\textregistered} bricks.

As introduced in Section~\ref{sec:prelim}, crisscross polymerization can be used to construct stable DNA nanostructures, including DNA-tiles.
Figure~\ref{fig:tiles}~(VI) illustrates how crisscross slats can be used to form DNA-tiles~\cite{drake2026slats}.
These slats are highly stable, which enables them to assemble into big structures relative to the nanoscale~\cite{minev2021crisscross}.
Figure~\ref{fig:tiles}~(VI)~(a) shows the mathematical representation of a single tile with glues on each side.
The northern and western glues are output glues, while the southern and eastern glues are input glues.
All glues have a strength of 1.
This tile can be represented by a set of slats divided into functional groups, as depicted in Figure~\ref{fig:tiles}~(VI)~(b).
Figure~\ref{fig:tiles}~(VI)~(c) shows how these slat groups assemble into the original tile.
The colored regions indicate different functions within the tile, including the representation of its output glues, as illustrated in Figure~\ref{fig:tiles}~(VI)~(d).
Figure~\ref{fig:tiles}~(VI)~(e) shows the mathematical representation of a tile assembly with two candidates that could attach at the same position.
The tile named $t_X$ can attach through a northern glue with strength 2, while the tile named $t_Y$ requires cooperative binding through western and southern glues of strength 1 each.
Figure~\ref{fig:tiles}~(VI)~(f) depicts the corresponding slat-based representation.
The tiles are color coded to indicate which slats belong to which tile.
The slats of t$_\text{X}$ and t$_\text{Y}$ are colored in magenta with indications of which regions of the slats can bind to the sticky ends of the other tiles.
Slats can be implemented using individual ssDNA strands or more complex DNA origami structures~\cite{wintersinger2023crisscross}.

The previously discussed candidates represent established and particularly suitable implementations of DNA tiles, but the concept is not limited to these structures.
DNA as a building material can be used to form a wide variety of nanostructures.
In principle, many DNA nanostructures can be adapted as DNA tiles as long as they provide sticky ends that enable assembly into larger and more complex structures.
Song et al.~show a collection of DNA based nanostructures in the medical context~\cite{song2017nanostructures}.
Most of the introduced structures can be adapted as tiles, but complexity and costs might be a constraining factor.
In the medical context, DNA hydrogels are an interesting implementation of DNA-tiles for controlled drug delivery and diagnostics~\cite{song2017nanostructures}.
Some DNA hydrogels like the enzyme-catalyzed assembly from~\cite{um2006enzyme} or the stimuli-responsive DNA hydrogel from~\cite{li2015hydrogel} might be adaptable to tiles as well.
Another complex nanostructure designed for the medical field is the nanoflower~\cite{zhu2013flower}.
It could be adapted as a tile as well, but these DNA nanostructures with additional functionalities are not further considered out of complexity reasons for the minimal proof-of-concept in this work.

As DNA strands are not solid and can deform naturally, we assume that minor changes in the design of tiles have little impact on the resulting structure.
For example, in this abstraction, a DX-tile with one sticky end on each side is treated equivalently to a DX-tile with two sticky ends on the left and two on the right.
With this assumption even more complex structures can be modeled with just their sticky ends as a deciding factor.
Something like this needs to be further researched in extensive wet-lab experiments and has no further relevancy for this paper.
For the purpose of this work, we therefore focus on which sticky ends can bind to each other rather than on the exact spatial geometry of the tile.
We assume that moderate deformation of a DNA-tile does not invalidate this abstraction as long as the intended sticky-end interactions remain possible.

% green DCD8BC (5| 22-25)     \score{DCD8BC}{5}
% blue CDDBD8 (4 | 18-21)     \score{CDDBD8}{4}
% yellow ECDA88 (3 | 13-17)   \score{ECDA88}{3}
% orange F0CDB2 (2 | 9-12)    \score{F0CDB2}{2}
% red E8C0AD (1 | 5-8)        \score{E8C0AD}{1}

\begin{table}
  \centering
  \caption{Component comparison heatmap for tiles in a minimal DNA-based nanonetwork PoC. The five parameters Amount of Subcomponents (AoS), Amount of Operations (AoO), Compatibility with DNA (CwD), Level of Stability (LoS) and Cost of Components (CoC) are scored 1--5.}
  \label{tab:component-tile-heatmap}
  \begin{tabular}{lrrrrrr}
    \toprule
    Component           & Score               & AoS                & AoO                & CwD                & LoS                & CoC                \\
    \midrule
    SST/DNA bricks      & \score{DCD8BC}{22}  & \score{DCD8BC}{5}  & \score{CDDBD8}{4}  & \score{DCD8BC}{5}  & \score{ECDA88}{3*} & \score{DCD8BC}{5}  \\
    Holliday junction   & \score{CDDBD8}{21}  & \score{CDDBD8}{4}  & \score{CDDBD8}{4}  & \score{DCD8BC}{5}  & \score{ECDA88}{3}  & \score{DCD8BC}{5}  \\
    DX-tile             & \score{CDDBD8}{21}  & \score{CDDBD8}{4}  & \score{CDDBD8}{4}  & \score{DCD8BC}{5}  & \score{CDDBD8}{4}  & \score{CDDBD8}{4}  \\
    PX-tile             & \score{CDDBD8}{20}  & \score{CDDBD8}{4}  & \score{CDDBD8}{4}  & \score{DCD8BC}{5}  & \score{ECDA88}{3}  & \score{CDDBD8}{4}  \\
    JX$_1$/JX$_2$-tile  & \score{CDDBD8}{20}  & \score{CDDBD8}{4}  & \score{CDDBD8}{4}  & \score{DCD8BC}{5}  & \score{ECDA88}{3}  & \score{CDDBD8}{4}  \\
    TX-tile             & \score{CDDBD8}{20}  & \score{CDDBD8}{4}  & \score{CDDBD8}{4}  & \score{DCD8BC}{5}  & \score{CDDBD8}{4}  & \score{ECDA88}{3}  \\
    three-helix bundle  & \score{CDDBD8}{20}  & \score{CDDBD8}{4}  & \score{CDDBD8}{4}  & \score{DCD8BC}{5}  & \score{CDDBD8}{4}  & \score{ECDA88}{3}  \\
    five-arm junction   & \score{CDDBD8}{20}  & \score{CDDBD8}{4}  & \score{CDDBD8}{4}  & \score{DCD8BC}{5}  & \score{ECDA88}{3}  & \score{CDDBD8}{4}  \\
    six-arm junction    & \score{CDDBD8}{19}  & \score{ECDA88}{3}  & \score{CDDBD8}{4}  & \score{DCD8BC}{5}  & \score{ECDA88}{3}  & \score{CDDBD8}{4}  \\
    eight-arm junction  & \score{CDDBD8}{19}  & \score{ECDA88}{3}  & \score{CDDBD8}{4}  & \score{DCD8BC}{5}  & \score{ECDA88}{3}  & \score{CDDBD8}{4}  \\
    six-helix bundle    & \score{CDDBD8}{18}  & \score{ECDA88}{3}  & \score{CDDBD8}{4}  & \score{DCD8BC}{5}  & \score{CDDBD8}{4}  & \score{F0CDB2}{2}  \\
    twelve-arm junction & \score{ECDA88}{17}  & \score{F0CDB2}{2}  & \score{CDDBD8}{4}  & \score{DCD8BC}{5}  & \score{ECDA88}{3}  & \score{ECDA88}{3}  \\
    DNA origami tile    & \score{ECDA88}{15}  & \score{E8C0AD}{1}  & \score{CDDBD8}{4}  & \score{DCD8BC}{5}  & \score{CDDBD8}{4}  & \score{E8C0AD}{1}  \\
    origami crisscross  & \score{ECDA88}{15*} & \score{E8C0AD}{1*} & \score{CDDBD8}{4*} & \score{DCD8BC}{5*} & \score{CDDBD8}{4*} & \score{E8C0AD}{1*} \\
    ssDNA crisscross    & \score{ECDA88}{14*} & \score{E8C0AD}{1*} & \score{CDDBD8}{4*} & \score{DCD8BC}{5*} & \score{ECDA88}{3*} & \score{E8C0AD}{1*} \\
    \bottomrule
    \footnotesize *low evidence
  \end{tabular}
\end{table}

With all these different implementations of DNA-tiles, we can compare them according to the parameters defined in the beginning of this section.
All scores discussed in the following paragraph, can be viewed in Table~\ref{tab:component-tile-heatmap}.
Looking at Holliday Junctions and multi-arm adaptations of Holliday junctions first, the amount of subcomponents depends on the number of arms.
The four-arm Holliday junction needs four oligonucleotides, the twelve-arm junction needs twelve oligonucleotides, accordingly.
The AoS score of DNA junctions therefore varies depending on their number of arms.
The amount of operations needed for the construction of junctions is constant.
One annealing operation is necessary to break unwanted bonds and assemble the up to twelve oligos into their predesigned junctions.
Holliday junctions are DNA-native and therefore fully compatible with other DNA structures.
Shah et al. discuss the stability of four-arm junctions and define $T_m$ at \SIrange{58}{72}{\degreeCelsius}~\cite{shah1994activation}.
The half-life $t_{50}$ and $T_m$ for all multi-armed junctions are not mentioned further.
Wang et al. discuss the stability requirements of different multi-arm junctions, including five- and six-arm junctions in~\cite{wang1991fivearm} and eight- and twelve-arm junctions in~\cite{wang2007eightarm}.
They come to the conclusion, that four-, five-, six- and eight-arm junctions need strands with 32 bases to be stable and twelve-arm junctions need strands with 48 bases for the same result.
The difference in stability is attributed to the friction in the center of the junctions.
Based on the lower bound of the reported melting temperatures for four-arm junctions and the structural requirements of other multi-arm junctions, we assign a LoS score of 3 to all considered junctions.
For the pricing of the junctions, we use the smallest possible tube from IDT, the \SI{25}{\nano\mole} DNA Oligo.
With €0,33 per base we estimate, the pricing for the oligonucleotides for all junctions as: €42,24 for four-arm junctions, €52,80 for five-arm junctions, €63,36 for six-arm junctions, €84,48 for eight-arm junctions, and €190,08 for twelve-arm junctions~\cite{IDT2025}.

The multi-crossover-tiles need a constant number of subcomponents.
DX-, PX, JX$_1$, JX$_2$, and TX-tiles are all composed of four ssDNAs~\cite{chandrasekaran2020px, labean2000txtiles}.
There are different adaptations of those design principles with more strands, for example implementations of TX-tiles with six strands are commonly used~\cite{zhang2013integrating, liu2004nanotubes}.
Just as with Holliday Junctions and multi-arm tiles, the amount of operations for multi-crossover-tiles is constant, and the compatibility is DNA native.
Chandrasekaran et al.~\cite{chandrasekaran2020px} analyze the stability of DX-, PX-, and JX-tiles.
They report melting temperatures of approximately \SI{60}{\degreeCelsius} for DX-tiles and \SI{55}{\degreeCelsius} for PX- and JX-tiles.
All three structures are stable for at least 24 hours at \SI{37}{\degreeCelsius}, resulting in a LoS score of 4 for DX-tiles.
Due to the lower melting temperature of PX- and JX-tiles, we assign them a LoS score of 3.
However, their resistance to degradation in biological fluids differs substantially, which is not included in our scoring, as we focus on experiments under controlled in vitro buffer conditions.
LaBean et al.~\cite{labean2000txtiles} discuss the stability of TX-tiles.
$T_m$ is monitored to be between \SIrange{65.2}{73.0}{\degreeCelsius} for TX-tiles, $t_{50}$ is not defined.
We estimate the cost of DX-tiles from the designs of Fu and Seeman~\cite{fu1993dx}.
The four strands needed for the construction have 38 bases, exactly the same as the PX-tile design by Shen et al.~\cite{shen2004px}.
JX$_1$ and JX$_2$ are analogous to PX-tiles, as they are topoisomers.
LaBean et al. designed the TX-Tile with two strands with 52 bases and two strands with 72 bases~\cite{labean2000txtiles}.
The costs for the multi-crossover-tiles are according to IDT with the smallest possible tubes €50,16 for DX-, PX-, JX$_1$- and JX$_2$-tiles~\cite{IDT2025}.
The 72-base strands require a larger minimum order quantity than the 52-base strands, resulting in a total estimated cost of €117.84 for the TX-tile~\cite{IDT2025}.

Analog to the design principle in Figure~\ref{fig:tiles} (II) b), we assume that the three-helix bundle and the six-helix bundle are build from TX-Tiles.
Accordingly, the score of the three-helix bundle is equivalent to the score of TX-tiles.
We assume that the six-helix bundle has the same amount of operations, DNA compatibility and level of stability, but double the amount of subcomponents and therefore costs.
The amount of subcomponents of SSTs is one and the amount of operations is one for an initial annealing process to denature unwanted bounds.
SSTs are natively compatible with DNA.
Wickstrom~\cite{wickstrom1986oligodeoxynucleotide} tested and discussed the stability of single-stranded DNA in different media.
The fastest degradation of the oligonucleotides is \SI{15}{\minute} in bovine calf serum.
Crine and Verly report a spontaneous strand-break rate of 0.009 breaks per hour for a DNA strand with 39,937 bases at \SI{37}{\degreeCelsius} and pH 7.21~\cite{crine1976degradation}.
Interpreting strand breaks as a Poisson process yields $t_{50}\approx77$ hours for the full-length strand.
Assuming that the break rate scales linearly with strand length, this corresponds to $t_{50}\approx$ 4--35 years for oligonucleotides with 90--10 bases.
As this estimate involves a substantial extrapolation to much shorter DNA strands and does not account for sequence- or environment-specific degradation, we assign SSTs a conservative LoS score of 3 with low evidence.
Wei et al.~use SSTs for their work and designed them with 42 bases, to have four binding domains~\cite{wei2012sst}.
The cost of one oligonucleotide from IDT with 42 bases in a \SI{25}{\nano\mole} tube, is €13,86.

The DNA origami design of Rothemund for tiles consists of one scaffold strand and over 200 short staple strands, for the amount of subcomponents~\cite{rothemund2006origami}.
The main principle comes down to a one-pot annealing process, but often another purification step is necessary~\cite{baptist2023lyophilization}.
We assume 1--2 operations because of that.
The compatibility to DNA is native, just as every other tile before.
Baptist and Heuer-Jungemann show that DNA origami structures are stable for at least four weeks in a salt buffer at room temperature~\cite{baptist2023lyophilization}.
Hanke et al. test melting temperatures of DNA origami structures with $T_m\geq\SI{60}{\degreeCelsius}$~\cite{hanke2023ionicstrength}, resulting in a LoS score of 4.
We estimate the cost of a DNA origami tile from the price of 200 oligonucleotides with 32 bases from IDT with approximately €2112 and the price of the scaffold strand from Tilibit with €112~\cite{IDT2025, TILIBIT2026}.

Crisscross polymerization uses components, that were already scored.
The concepts of crisscross polymerization tiles from Drake et al.~\cite{drake2026slats} that were introduced in Figure~\ref{fig:tiles} (VI) are not experimentally validated yet, which makes scoring more difficult.
We estimate the scores based on related experimental systems from Wintersinger et al.~\cite{wintersinger2023crisscross} and Minev et al.~\cite{minev2021crisscross}.
Both papers test crisscross polymerization for DNA structures but not for DNA-tiles in particular.
We therefore mark all scores for crisscross polymerization as low evidence in Table~\ref{tab:component-tile-heatmap}.
For the AoS, we use a conservative estimate of more than 50 subcomponents.
The number of operations depends on the type of slat.
For ssDNA-based crisscross polymerization, the structure can be assembled in a single step, whereas DNA-origami-based slats require at least one additional fabrication step before polymerization.
Both results in a AoO score of 4.
For the LoS score, we apply the same scores from ssDNA- and DNA-origami-tiles, as the slats are based on these designs.
The component cost is estimated to exceed €500 for both ssDNA- and DNA-origami-based crisscross polymerization because of the potentially large number of required slats.
Because several of these values rely on conservative estimates rather than directly reported experimental data, all scores for crisscross polymerization as a tile are marked as low evidence in Table~\ref{tab:component-tile-heatmap}.

According to the introduced metric, the best implementation of DNA-tiles based on Table~\ref{tab:component-tile-heatmap} for a minimal proof-of-concept experiment of DNA-based nanonetworks is the single-stranded tile.
Overall, the evaluated tile implementations achieve relatively high scores, as the analysis deliberately focuses on established DNA nanostructures that have already been experimentally demonstrated and could, in principle, all be used for such an implementation.
The main concern of SSTs is their stability in media that is not ideal for DNA.
If the stability is deemed to be the most important factor, DX-tiles as a relatively cheap implementation of DNA-tiles could be an alternative to SSTs.
If cost is not a limiting factor, DNA origami is an alternative with one of the highest stability scores in our evaluation.
We assume, that the stability of SSTs suffices for a minimal proof-of-concept wet-lab experiment.
SSTs further minimize the possible points of failure, due to their low complexity.
Also, the low costs of buying single oligonucleotides as tiles make the experiment more feasible and repeatable.
Crisscross polymerization has low scores based on the experimental data available and our conservative estimates.
If the concepts and simulations of Drake et al.~\cite{drake2026slats} can be experimentally validated, we expect that ssDNA crisscross polymerization tiles will be a good alternative to SSTs.

%%%%%%%%%%%%%%%%%%%%%%%%%%%%%%%%%%%%%%%%%%%%%%%%%%%%%%%%%%%%%%%%%%%%%%%%%%%%%%%%%%%%%%%%%%%%%%%
%%% SOTA BOXES %%%%%%%%%%%%%%%%%%%%%%%%%%%%%%%%%%%%%%%%%%%%%%%%%%%%%%%%%%%%%%%%%%%%%%%%%%%%%%%%
%%%%%%%%%%%%%%%%%%%%%%%%%%%%%%%%%%%%%%%%%%%%%%%%%%%%%%%%%%%%%%%%%%%%%%%%%%%%%%%%%%%%%%%%%%%%%%%

\subsection{Nanocarriers}
\label{subsec:nanocarriers}

\begin{figure*}
  \centering
  \includegraphics[width=\textwidth]{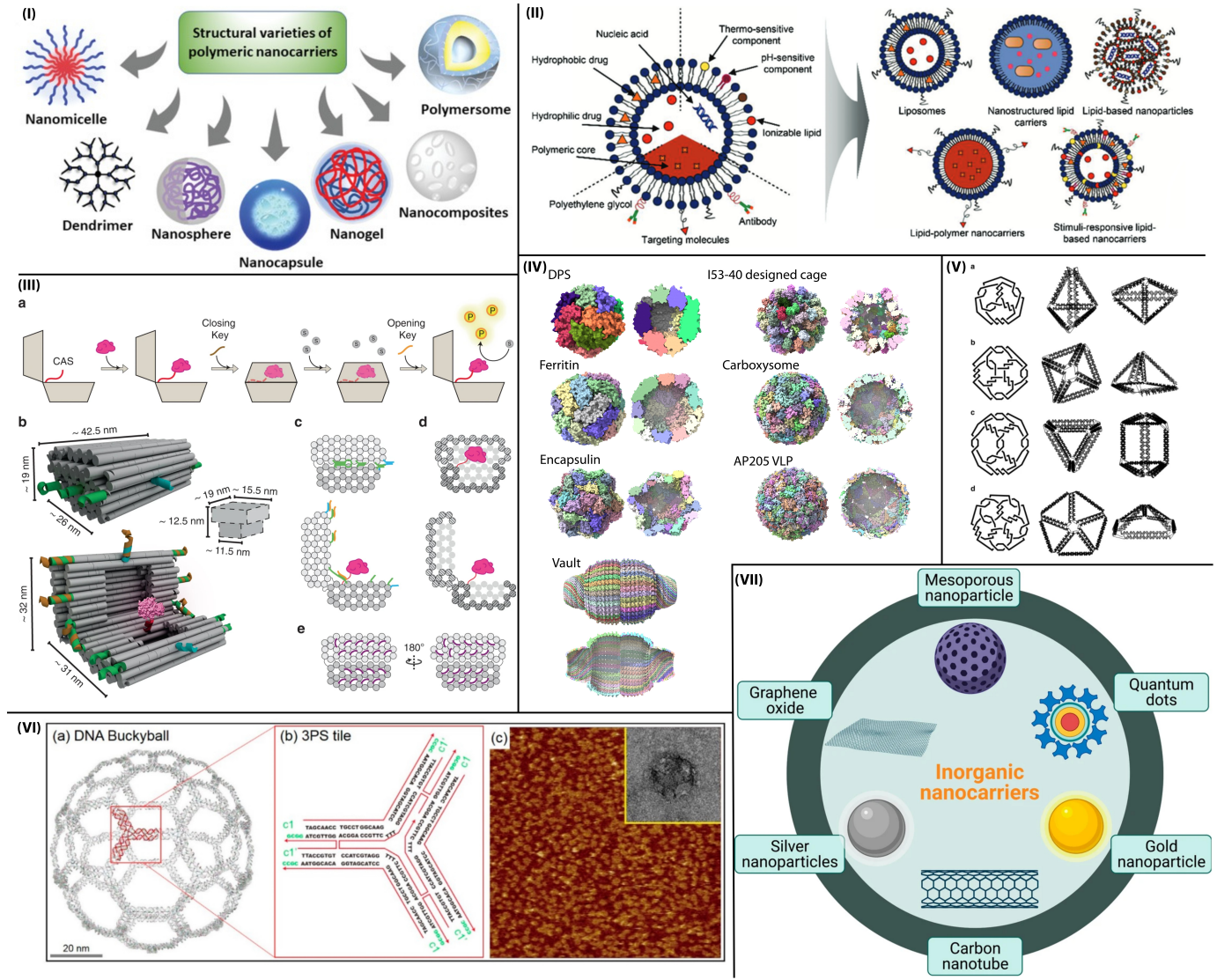}
  \caption{A collection of different types of nanocarriers. (I) Visualization of different polymeric nanocarriers, reproduced from~\cite{de2022polymeric}, licensed under CC BY 4.0. (II) Schematic of possible functionalities and variations of lipid-based nanocarriers on the left and five examples on the right, reproduced from~\cite{zhao2022lipid}, licensed under CC BY 4.0. (III) Concept, model, and schematic representation of a DNA-origami-based nanocarrier, reproduced from~\cite{grossi2017origamibox}, licensed under CC BY 4.0. (a) illustrates the concept and functionality of the nanocarrier, while (b) presents its 3D model, including its dimensions. (c), (d), and (e) depict the nanocarrier's structure from different perspectives and in different planes. (IV) Different protein-based nanocarriers visualized with ChimeraX using structures from the Protein Data Bank (PDB). The complete structures are shown on the left and cross-sectional views on the right. Depictions are not to scale; the selection is based on~\cite{sheng2025nanocages}. (V) Different shapes of simple DNA cages, reproduced from~\cite{qi2018sstboxes}, licensed under CC BY 4.0. (VI) DNA-tile-based cage showing (a) the model, (b) the schematic of a single DNA tile, and (c) AFM images of the cages, reproduced from~\cite{jo2018tileboxes}, licensed under CC BY 4.0. (VII) Overview of inorganic nanocarrier types, reproduced from~\cite{kaushik2022inorganic}, licensed under CC BY 4.0.}
  \label{fig:nanocarriers}
\end{figure*}

As another core component of the proposed DNA-based nanonetwork, nanocarriers serve as the physical implementation of the tile-releasing nanodevices shown in Figure~\ref{fig:nanonetwork}.
Their function is to carry DNA-tiles and release them in response to defined events like biomarker detection.
Since nanocarriers can differ substantially in material and structure, this subsection examines selected design candidates and evaluates their suitability for the proof-of-concept implementation.
We pay particular attention to their ability to hold and release cargo.
Hofmann et al. review a big variety of nanocarriers for analytical applications~\cite{hofmann2019nanocontainers}.
Many reviews view a certain class or type of nanocarrier and discuss multiple instances or implementations of the same type~\cite{lee2016bioengineered, yin2025cavity, bhaskar2017protein, wang2024nanocages, pottanam_chali2020polymer}.

Arora et al.~\cite{arora2025nanocontainers} classify bio-nanocontainers into five different classes with four subclasses attributed to them.
The literature uses terms such as \textit{nanobox}, \textit{nanocarrier}, \textit{nanocage} and \textit{nanocontainer} for partially overlapping classes of cargo holding nanostructures.
In this paper, we use \textit{nanocarrier} as the general term for those components.
We retain source-specific terminology when discussing individual designs.
Table~\ref{tab:nanocarrier-classes} show the classes and their attributed subclasses.
Every nanocarrier can be described by one or more subclasses.
Many nanocarriers are designed to release their payload by degradation of the nanocarriers in the destined medium or cell, based on stimuli.
The following paragraphs first provide an overview of the considered nanocarrier implementations and their relevant properties before evaluating them according to the defined metrics.

\begin{table*}
  \caption{Different classes and their subclasses of bio-nanocontainers according to Arora et al.~\cite{arora2025nanocontainers}}
  \begin{tabular}{ll}
    \hline
    Class                & Subclasses                                                                   \\\hline
    Based on composition & lipid-based, polymer-based, protein-based, inorganic-based                   \\
    Based on origin      & natural, synthetic, semisynthetic, biomimetic                                \\
    Based on structure   & vesicular, capsular, core-shell, self-assembled                              \\
    Based on function    & drug delivery, gene delivery, imaging/diagnostics, vaccination/immunotherapy \\
    Based on stimuli     & pH-sensitive, thermosensitive, light-sensitive, redox-sensitive              \\\hline
  \end{tabular}
  \label{tab:nanocarrier-classes}
\end{table*}

\textit{Polymeric nanocarriers} (PNCs) are a wide field of natural and synthetic nanoparticles that are discussed in many reviews, mostly in a medical context~\cite{jan2024polymer, romero-ben2025polymer, ma2021polymer, song2023polymeric}.
PNC is a collective term that encompasses many different nanocontainers.
Figure~\ref{fig:nanocarriers} (I)~\cite{de2022polymeric} shows a collection of different PNCs.
Nanomicelles consist of self-assembled amphiphilic copolymers, that have a hydrophobic and a hydrophilic block.
Based on the type of micelle, they can have a hydrophilic core with a hydrophobic corona, a hydrophobic core with a hydrophilic corona or a mixed composition of copolymers~\cite{de2022polymeric}.
Dendrimers are hyperbranched polymers with many different types and structures~\cite{rajabi2025polymer}.
They are designed on atomic or molecular level and highly diverse, which makes them very versatile and small even for nano scale.
Nanospheres, Nanocapsules and Polymersomes are polymeric structures with different copolymer unimer compositions~\cite{letchford2007polymer}.
Nanogel is the general term for all aqueous dispersions of cross-linked polymeric particles~\cite{kubeil2024nanogels}.
The most common nanogels are hydrogels, which consist of cross-linked or self-organizes hydrophilic polymers.
They swell when containing water molecules.

\textit{Nano Lipid-based carriers} (NLBCs) are another type of nanocarrier, that is widely used in the medical context~\cite{kumar2022lipid}.
NLBCs are mainly used in the medical context due to their biocompatibility and capacity to carry hydrophobic medical drugs~\cite{attri2024liposome}.
As shown in Figure~\ref{fig:nanocarriers} (II) NLBCs are vesicles with many different combinable functionalities~\cite{zhao2022lipid}.
Liposomes are vesicles typically build from a phospholipid bilayer~\cite{attri2024liposome}.
They are amphiphilic, which means that they have hydrophilic and hydrophobic properties combined in one structure.
By that, they can carry a hydrophobic molecule in between the lipid bilayer and a hydrophilic molecule in its core.
This can be seen in Figure~\ref{fig:nanocarriers} (II) with a hydrophilic drug marked as orange triangles and with a hydrophobic drug marked as red dots.
Liposomes can be loaded passively or actively~\cite{akbarzadeh2013lipo}.
For both mechanisms, a lot of different techniques are used, depending on the type of cargo.
Generally, passive loading incorporates cargo during liposome formation, while active loading introduces cargo into preformed liposomes, often using transmembrane pH or ion gradients.
\textit{Nanostructured lipid carriers} (NLCs) are also known as second-generation NLBCs~\cite{kumar2022lipid}.
They use both a solid and a liquid lipid to increase the drug loading efficiency.
Figure~\ref{fig:nanocarriers} (II) shows the solid lipid with the copper colored capsules inside the NLCs.
The solid-lipid and liquid-lipid combination turns into a disordered crystal structure that can hold more cargo than normal liposomes.
Nanoemulsions, Niosomes, ionizable lipid nanoparticles, solid lipid nanoparticles and lipid-based nanomicelles are all different types of NLBCs with different characteristics.
A combination of polymer and lipid-based nanoparticles are lipid polymer hybrid nanoparticles.
They were designed to overcome the limitations of liposomes, which are uncontrolled drug release, instability and insufficient drug loading~\cite{zhao2022lipid}.
Commonly, lipid-based nanoparticles use \textit{polyethylene glycol} (PEG) to increase the lipid-based carriers stability.
PEG is mainly used to increase the lifetime of the carriers in blood~\cite{kumar2022lipid}.
Lipid-based nanoparticles can also utilize antibodies or other types of targeting molecules as their targeting mechanism~\cite{zhao2022lipid}.
The lipid design can additionally incorporate responsiveness to multiple different stimuli, as illustrated in Figure~\ref{fig:nanocarriers} (II).

The native design principle of DNA nanocarriers addresses a major concern of our proof-of-concept experiment: DNA compatibility.
A comprehensive and detailed review about them was written by Chandrasekaran and Levchenko~\cite{chandrasekaran2016dna_nanocages}, which lays the foundation for the following paragraph.
They distinguish four broad strategies for constructing DNA cages: modular assembly, hierarchical assembly of DNA tiles, one-pot assembly of individual DNA strands, and DNA origami.
Depending on the cage and cargo, specific binding sites can be incorporated into the structure to position DNA strands, proteins, nanoparticles, or other molecules inside the carrier.
The first strategy of DNA origami uses the design principle to design a hollow DNA structure.
DNA origami nanocarriers are of particular interest for biomedical applications because their programmable structures enable precise cargo loading, molecular functionalization and stimuli-responsive release mechanisms~\cite{rezaee2026origami}.
An example can be seen in Figure~\ref{fig:nanocarriers}~(III).
As introduced in the preliminary work, the DNA origami is constructed from a ssDNA-scaffold and many short staple strands.
Figure~\ref{fig:nanocarriers}~(III)~(a) shows the concept of holding cargo, closing the cage, opening the cage , and releasing the cargo, while (b--e) show the DNA origami design for such a DNA cage.
This design strategy enables comparatively large and precisely addressable DNA cages with defined internal cavities.
A second strategy is the one-pot assembly of individual DNA strands.
In this approach, several designed oligonucleotides are annealed together in a single assembly process to directly form a three-dimensional cage.
Figure~\ref{fig:nanocarriers}~(V)~(a--d) shows examples of such structures.
Another design principle is DNA-tile-based cages.
In this case, tiles self-assemble and then form the desired shape by connecting with their sticky ends.
Figure~\ref{fig:nanocarriers}~(VI) shows an exemplary DNA cage built from DNA tiles.
In (a), three-arm tiles form the cage, while (b) depicts the individual tiles in more detail.
Figure~\ref{fig:nanocarriers}~(VI)~(c) shows AFM images of such nanocarriers.
The fourth approach is modular assembly.
For this strategy, two or more components representing parts of the final nanocage are assembled separately before being combined into the final three-dimensional structure.
The individual modules are designed with compatible interaction sites, such as complementary sticky ends, that enable self-assembly.
Cargo loading is not restricted to one mechanism across these different DNA nanocages.
Cargo can, for example, be tethered to defined positions within the cage before assembly, captured during cage formation, or loaded into structures that can subsequently be opened and closed.
Depending on the loading strategy, unbound cargo and excess DNA components have to be removed by an additional purification step.
Sequence-specific hybridization sites, aptamers, or other molecular recognition mechanisms can additionally be incorporated to retain or position cargo within the nanocage.

Protein-based nanocarriers represent another common class of nanocarriers.
A detailed look on those types of nanocarriers was given by Sheng et al.~\cite{sheng2025nanocages}.
As seen in Figure~\ref{fig:nanocarriers} (IV) there are many protein-based nanocontainers of different forms and sizes.
They are all built by spontaneous self-assembly and can be found in the \href{https://www.rcsb.org/}{RCSB Protein Data Bank}\footnote{\url{https://www.rcsb.org/}}.
From top to bottom and left to right in Figure~\ref{fig:nanocarriers} (IV), the first shown protein-based carrier is called  \textit{DNA-binding proteins from the starved cells} (DPS, PDB code: \href{https://www.rcsb.org/structure/6LKP}{6LKP}).
This carrier type self-assembles and can be found in various organisms as it is generally present in bacteria to shell the bacterial cells from oxidative damages.
The \textit{Ferritin} (PDB code: \href{https://www.rcsb.org/structure/2X17}{2X17}) protein-based carrier is just as common, as its use is to store iron particles in a non-toxic form inside bio organisms.
Ferritin is of special interest for targeted drug delivery, due to its structure, targeting ability and biocompatibility~\cite{zhang2021ferritin}.
\textit{Encapsulin} (PDB code: \href{https://www.rcsb.org/structure/6NJ8}{6NJ8}) is widely spread in archaea and bacteria and can occur in different forms depending on its cargo.
\textit{Vault} (PDB code: \href{https://www.rcsb.org/structure/4V60}{4V60}).
It is a structural protein and among the largest cytoplasmic ribonucleoprotein particles.
First on the right side in Figure~\ref{fig:nanocarriers} (IV), the \textit{I53-40 designed cage} (PDB code: \href{https://www.rcsb.org/structure/5IM5}{5IM5}) is a computationally designed protein nanocage.
In contrast to the previously described naturally occurring protein carriers, it was specifically designed to self-assemble from two different protein components into an icosahedral cage.
Its controlled structure and assembly make it particularly interesting as a customizable carrier for molecular cargo.
Next are \textit{Carboxysomes} (PDB code: \href{https://www.rcsb.org/structure/9F0H}{9F0H}).
They encapsulate $\mathrm{RuBisCo}$ and thereby help to concentrate $\mathrm{CO_2}$ around the enzyme.
The last exemplary protein-based nanocarrier in Figure~\ref{fig:nanocarriers} (IV) is the \textit{AP205 virus-like particle (VLP)} (PDB code: \href{https://www.rcsb.org/structure/5LQP}{5LQP}).
It is a virus-like particle formed by the self-assembly of coat proteins from the bacteriophage AP205.
As they form a stable protein shell without requiring an infectious viral genome, AP205 VLPs are suitable as carriers for molecular cargo and as platforms for the presentation of molecules on their surface.
While protein-based nanocarriers are powerful and highly stable, their architectures are largely predetermined.
Any functional modification must remain compatible to their original design principle of folding and self-assembly.

Some of the already mentioned nanocarriers are inorganic.
This category covers a broad range of materials with different optical, magnetic and chemical properties.
For our experimental setup, many inorganic nanocarriers are less suitable, as their biological compatibility and toxicity can strongly depend on their material, size, surface chemistry, and functionalization~\cite{luther2020inorganic,unnikrishnan2023inorganic}.
Not all inorganic nanocarriers are incompatible with biological systems, and several materials have been extensively investigated for biomedical applications.
Figure~\ref{fig:nanocarriers}~(VII) shows a schematic of six materials commonly used for inorganic nanocarriers in biological contexts.
\textit{Mesoporous silica nanoparticles} (MSNs) are the most commonly used silica-based nanocarriers.
MSNs have a honeycomb-like structure with hollow channels and pores that can hold cargo\cite{luther2020inorganic, valletregi2018msn}.
They can be constructed in many different shapes and sizes and in combination with other nanocarrier technologies, which makes them very versatile~\cite{narayan2018msnp, noureddine2023msnp}.
\textit{Quantum dots} (QDs) are nanoscale semiconductor crystals with characteristic optical properties.
Their surfaces can be functionalized with drugs and targeting ligands, while their fluorescence additionally allows the carrier and its distribution to be tracked~\cite{yong2012quantumdots}.
However, especially conventional quantum dots containing heavy metals such as cadmium raise toxicity concerns, which limits their applicability in biological systems~\cite{valizadeh2012quantumdots}.
Gold nanoparticles are another type of nanocarrier widely investigated for biomedical applications due to their functionalization capabilities and potentially low toxicity since the late 1990s~\cite{luther2020inorganic}.
With many different strategies the properties of gold can be functionalized with a variety of ligand species and are therefore highly targeted carriers.
In addition to carrying molecular cargo, gold nanoparticles can convert absorbed light into heat and can therefore also be used for localized photothermal treatment~\cite{mioc2019gold}.
\textit{Carbon nanotubes} (CNTs) are cylindrical carbon-based nanostructures with a large surface area.
Cargo molecules can be attached to their surface or associated through non-covalent interactions, while additional surface functionalization can improve solubility and targeting~\cite{wong2013cnt}.
Their biological compatibility, however, strongly depends on properties such as size, structure, purity, and surface functionalization~\cite{debnath2021carbon}.
\textit{Silver nanoparticles} (AgNPs) combine their function as carrier platforms with an intrinsic antimicrobial activity~\cite{burdusel2018silver}.
Their surfaces can be functionalized or combined with therapeutic molecules for drug delivery applications.
However, their biological effects and toxicity strongly depend on properties such as particle size, concentration, and surface chemistry, which have to be considered when using them as carriers~\cite{burdusel2018silver,unnikrishnan2023inorganic}.
\textit{Graphene oxide} (GO) is a two-dimensional carbon-based nanomaterial with a large surface area and numerous oxygen-containing functional groups.
These properties allow drugs and other biomolecules to be adsorbed onto or conjugated to its surface, making GO an interesting platform for drug delivery~\cite{liu2013graphene}.
Similar to CNTs, its surface can be further functionalized to modify its dispersibility, targeting and biological interactions~\cite{liu2013graphene,debnath2021carbon}.
Next to the shown examples there are other feasible inorganic nanocarriers.
Lanthanide \textit{upconversion nanoparticles} (UCNPs) are a class of nanomaterials, that use optical and photochemical properties to hold and release their cargo.
They have low toxicity which makes them good candidates for controlled and targeted drug delivery~\cite{luther2020inorganic}.
Iron oxide nanoparticles are biocompatible nanoparticles with magnetic properties with their three most common forms Fe$_3$O$_4$ Magnetite, $\alpha$-Fe$_2$O$_3$ Hematite and $\gamma$-Fe$_2$O$_3$ Maghemite in medical contexts~\cite{rahman2011iron}.
Other inorganic materials and nanoparticles can be used as nanocarriers such as zinc oxide, cerium oxide and hydroxyapatite~\cite{unnikrishnan2023inorganic}.

% green DCD8BC (5| 22-25)     \score{DCD8BC}{5}
% blue CDDBD8 (4 | 18-21)     \score{CDDBD8}{4}
% yellow ECDA88 (3 | 13-17)   \score{ECDA88}{3}
% orange F0CDB2 (2 | 9-12)    \score{F0CDB2}{2}
% red E8C0AD (1 | 5-8)        \score{E8C0AD}{1}
\begin{table}
  \centering
  \caption{Component comparison heatmap for nanocarriers in a minimal DNA-based nanonetwork PoC. The five parameters Amount of Subcomponents (AoS), Amount of Operations (AoO), Compatibility with DNA (CwD), Level of Stability (LoS) and Cost of Components (CoC) are scored 1--5.}
  \label{tab:component-nanocarrier-heatmap}
  \begin{tabular}{lrrrrrr}
    \toprule
    Component            & Score              & AoS               & AoO                & CwD                & LoS                & CoC               \\
    \midrule
    MSN                  & \score{CDDBD8}{20} & \score{DCD8BC}{5} & \score{CDDBD8}{4}  & \score{CDDBD8}{4}  & \score{DCD8BC}{5*} & \score{F0CDB2}{2} \\
    ssDNA cages          & \score{CDDBD8}{18} & \score{CDDBD8}{4} & \score{CDDBD8}{4}  & \score{DCD8BC}{5}  & \score{ECDA88}{3}  & \score{F0CDB2}{2} \\
    Polymersomes         & \score{CDDBD8}{18} & \score{DCD8BC}{5} & \score{F0CDB2}{2}  & \score{CDDBD8}{4*} & \score{DCD8BC}{5*} & \score{F0CDB2}{2} \\
    DNA-tile-based cages & \score{ECDA88}{17} & \score{F0CDB2}{2} & \score{CDDBD8}{4}  & \score{DCD8BC}{5}  & \score{CDDBD8}{4}  & \score{F0CDB2}{2} \\
    Modular DNA cages    & \score{ECDA88}{16} & \score{F0CDB2}{2} & \score{CDDBD8}{4}  & \score{DCD8BC}{5}  & \score{ECDA88}{3}  & \score{F0CDB2}{2} \\
    DNA-origami cages    & \score{ECDA88}{15} & \score{E8C0AD}{1} & \score{CDDBD8}{4}  & \score{DCD8BC}{5}  & \score{CDDBD8}{4}  & \score{E8C0AD}{1} \\
    Liposomes            & \score{ECDA88}{15} & \score{CDDBD8}{4} & \score{F0CDB2}{2}  & \score{ECDA88}{3}  & \score{ECDA88}{3*} & \score{ECDA88}{3} \\
    Encapsulin           & \score{ECDA88}{15} & \score{CDDBD8}{4} & \score{E8C0AD}{1*} & \score{F0CDB2}{2}  & \score{CDDBD8}{4*} & \score{CDDBD8}{4} \\
    \bottomrule
    \footnotesize *low evidence
  \end{tabular}
\end{table}

With so many different and highly specific nanocarriers, scoring every single one is beyond the scope of this work and our minimal proof-of-concept experiment.
Therefore, we select one representative for polymer-based, lipid-based, protein-based, and inorganic nanocarriers.
These representatives do not cover the full capabilities of their respective categories but provide suitable implementations for evaluating their applicability to our use case.
DNA-based nanocontainers are considered separately because of their native compatibility with the DNA tiles in our system.
Table~\ref{tab:component-nanocarrier-heatmap} displays the results.

We chose liposomes as the representative of lipid-based nanocarriers, that are mainly built from two subcomponents, cholesterol and phospholipids~\cite{akbarzadeh2013lipo}.
A third subcomponent might be necessary, depending on the loading technique.
We assume that some sort of positively charged lipid is needed to load DNA-tiles, as seen in the work of Barba et al. and the loading of Nucleic Acid Based Drugs~\cite{barba2019lipid}.
With three subcomponents we score the AoS with 4.
All methods used to prepare liposomes follow four basic steps~\cite{akbarzadeh2013lipo}.
Most loading techniques include one or two additional steps~\cite{akbarzadeh2013lipo}.
We conservatively assume six steps for the whole process, which results in an AoO score of 2.
The compatibility of lipid nanoparticles as nanocarriers for DNA is discussed in detail by Zhao and Huang~\cite{zhao2014lipid}.
They find that lipid-based nanoparticles as a whole can hold Gene material of nano- to micro-size with the stability and half-life of liposomes being a concern.
With the assumption of using PEGylated liposomes, with higher stability of carrier and payload, we give liposomes a score of 3 for their compatibility with DNA.
Jyothi et al. discuss the stability of liposomes in detail~\cite{jyothi2022liposome}.
They do not define $t_{50}$ or $T_m$ specifically, but from their very detailed analysis and suggestion of storing liposomes at \SIrange{2}{8}{\degreeCelsius}, we score the level of stability as 3.
The cost is scored with 3, because a single vial from the Sigma-Aldrich liposome kit costs €134,30~\cite{sigmaaldrich2026}.

As a representative of polymer-based nanocarriers, we consider polymersomes, which can be regarded as the polymeric counterpart of liposomes.
The amount of subcomponents for polymersomes can be held analogously to liposomes, but they are synthesized from only one block copolymer instead of cholesterol and phospholipids\cite{gomes2022poly}.
Therefore, the AoS score is 5 for one main building block and one additional targeting component used for the loading mechanism.
The amount of operations is similar to liposomes, which results in the same conservative score of 2.
Even though there is no specific research on their compatibility with DNA-tiles, we conclude the compatibility from the comparison of liposomes and polymersomes~\cite{rideau2018lipopoly}.
Polymersomes can significantly reduce cargo leakage due to their thicker and less permeable membranes.
For our concept, leaking nanocarriers need to be avoided, which increases the compatibility of polymersomes from our point of view.
A concrete specification of $t_{50}$ or $T_m$ is not possible for polymersomes, but from reviews about them a very high stability is their main purpose.
We therefore give polymersomes a score of 5 for the level of stability.
The cost of one pack of diblock copolymers used for polymersomes from Sigma-Aldrich is €460, which results in a CoC score of 2~\cite{sigmaaldrich2026}.

Encapsulin as the representative of protein-based nanocarriers self-assembles from one to three types of capsid proteins, depending on the type of protein used~\cite{giessen2016encapsulin}.
Additional targeting peptides need to be used for most types of cargo, which results in an AoS score of 4~\cite{kwon2025encapsulin}.
The experimental setup of functional, cargo-loaded encapsulin is complex and requires multi-step protocols that vary significantly based on the experimental context~\cite{sutter2008encapsulin,kwon2025encapsulin}.
Due to this high operational demand and procedural variability, we score the AoO of encapsulin with 1.
The compatibility of encapsulin regarding DNA is shortly discussed in~\cite{kwon2025encapsulin}.
By incorporating the nucleic acid-binding peptide Dps-N into the encapsulin shell, researchers demonstrated a loading mechanism for RNA that could potentially be adapted for DNA.
Due to the limited experimental evidence and the need for additional modifications, we conservatively score the CwD of encapsulin with 2.
As with other nanocarriers, the exact values for $t_{50}$ and $T_m$ are undefined for encapsulin, but experiments show a good stability in organic environments, which earns it a LoS score of 4~\cite{kwon2025encapsulin}.
Encapsulin plasmids cost about €80 in standard format from Addgene, which results in a CoC score of 4.

The amount of subcomponents for MSNs as the representative of inorganic nanocages can be generalized as the one specific silica nanoparticle.~\cite{narayan2018msnp}.
MSNs can be loaded after being fully synthesized with no further targeting molecule needed.
The AoS score therefore is 5.
The amount of operations is dependent on the loading method, which is either solvent-free or solvent-based~\cite{benko2025msnp}.
Solvent-free loading methods always need one single step of mixing the nanoparticles with their cargo.
Solvent-based loading methods take two steps:
The first step of mixing a solution of the cargo with the carrier and a second step of removing the solvent, that did not enter the pores of the MSNs.
Loading DNA into MSNs is possible without any additional modification, which results in an AoO score of 4.
The compatibility with DNA is given in the pore design of the MSN, with specific MSNs being able to hold and release nucleic acid molecules of different size and form~\cite{narayan2018msnp}.
MSNs are typically used for controlled cargo release over an extended period of time, instead of holding the cargo without leakage and releasing it on a specific trigger.
For that, additional steps are needed to functionalize and plug the pores of the MSN~\cite{slowing2007msnp}.
The compatibility with DNA is not native but can be achieved with little effort as explained, which results in a CwD score of 4.
The stability of MSNs can be scored with 5, as to the high temperature resistance, which is used to mix and load most MSNs~\cite{narayan2018msnp}.
Additionally, the biodegradability is discussed as a problem, because of its long process in biomedical applications~\cite{hu2021msnp}.
With adaptations this problem could be solved, as the complete excretion of MSNs out of the human body can take several weeks, while the biodegradation process takes several days~\cite{hu2021msnp}.
The cost of MSN is dependent on particle size, pore size and other additional characteristics.
A pack of powdered silica mesoporous SBA-15 with a pore size of \SI{12}{\nano\meter} from Sigma Aldrich costs €294, resulting in a CoC score of 2~\cite{sigmaaldrich2026}.

Lastly, we score all four DNA-based nanocontainer paradigms, as they are of native interest for DNA-based nanonetworks.
The modular self-assemblies, tile-based self-assemblies, single-stranded DNA self-assemblies and DNA origami-based self-assemblies of nanocontainers all consist of nucleic acids~\cite{chandrasekaran2016dna_nanocages}.
The compatibility therefore is just as all implementations of DNA-tiles in the last subsection native and therefore results in a CwD score of 5 for all four implementations.
The LoS of all types of DNA-based nanocontainers is equal to their DNA-tile counterpart from the previous subsection.
This results in a LoS score of 3 for single-stranded DNA self-assemblies and a LoS score of 4 for DNA origami-based self-assemblies.
The tile-based nanocontainers designed by Chandraskeran et al. follow the DX-tiles design paradigm and therefore gain the same LoS score of 4~\cite{chandrasekaran2016dna_nanocages}.
The modular nanocages have the same design principle as multiarmed junctions and therefore get a LoS score of 3.
We take the tetrahedral shape for the comparison of the amount of subcomponents, as it is a common DNA structure~\cite{zhang2020dna, cheng2022dna}.
The single-stranded tetrahedral nanocarriers need 4 strands and get an AoS score of 4.
For modular and DNA-tile-based nanocages, we assume four sequence-distinct structures.
This design aims to reduce unintended interactions and the formation of larger, uncontrolled assemblies.
For a conservative estimate of the complexity, we count all strands of the four building units as distinct molecular species.
Modular tetrahedral nanocontainers can be designed with four 3-armed junctions, that need three strands each, which results in twelve subcomponents and an AoS score of 2.
The tile-based self-assemblies need 4 DNA-tiles with six strands each, which gets an AoS score of 2 for the 24 subcomponents needed.
The DNA-origami-based nanocages need a lot more than 50 subcomponents, because of the design paradigm of DNA-origami and their very short staple strands, which gains an AoS score of 1.
The amount of operations is analog to their DNA-tile counterpart, just as the stability.
The scores therefore are 4 for single-stranded, modular, tile-based and DNA origami-based nanocontainers.
For the cost of the components, we take the tetrahedral shape as a ground truth again.
The nanocontainers need additional self-loops or other sticky ends to hold the DNA-tiles as their cargo.
We follow that the strands need to be up to 50\% longer for those sticky ends and self-loops to be stable.
The prices for DNA-tile-based and DNA-origami-based can be estimated from the prices of the DNA-tiles in the last subsection.
For that, we take the costs of the DNA-tiles from last subsection and multiply those by $1.5 \times 4 = 6$.
We get a CoC score of 1 for DNA-origami-based nanocontainers and a CoC score of 2 for DNA-tile-based nanocontainers.
The modular nanocontainers get a CoC score of 2 as well, because their design principle might be different to DNA-tile-based ones, but the amount of material is similar.
Single-stranded nanocontainers can be estimated as four strands of roughly 90 bases to form a stable tetrahedral shape with functional sticky ends and self-loops.
This results in a CoC score of 2 with an estimated price of €208.80 for the four strands from IDT~\cite{IDT2025}.

Based on the final scores found in Table~\ref{tab:component-nanocarrier-heatmap}, MSNs are the most fitting implementation for nanocarriers in a minimal proof-of-concept experiment for DNA-based nanonetworks.
MSNs are heavily researched and tested and have a high variability in size and opening mechanisms.
Concerns for MSNs are the increased complexity, the compatibility with DNA-tiles and the general biocompatibility.
If biocompatibility or compatibility with DNA is important, all four DNA-based nanocontainer implementations could provide more suitable nanocarriers.
Among them, ssDNA cages receive the highest score, but their small size could make DNA-origami cages more favorable, since the scoring does not consider size.

%%%%%%%%%%%%%%%%%%%%%%%%%%%%%%%%%%%%%%%%%%%%%%%%%%%%%%%%%%%%%%%%%%%%%%%%%%%%%%%%%%%%%%%%%%%%%%%
%%% SOTA SENSORS %%%%%%%%%%%%%%%%%%%%%%%%%%%%%%%%%%%%%%%%%%%%%%%%%%%%%%%%%%%%%%%%%%%%%%%%%%%%%%
%%%%%%%%%%%%%%%%%%%%%%%%%%%%%%%%%%%%%%%%%%%%%%%%%%%%%%%%%%%%%%%%%%%%%%%%%%%%%%%%%%%%%%%%%%%%%%%

\subsection{Biosensors}
\label{subsec:biosensors}

\begin{figure*}
  \centering
  \includegraphics[width=0.9\textwidth]{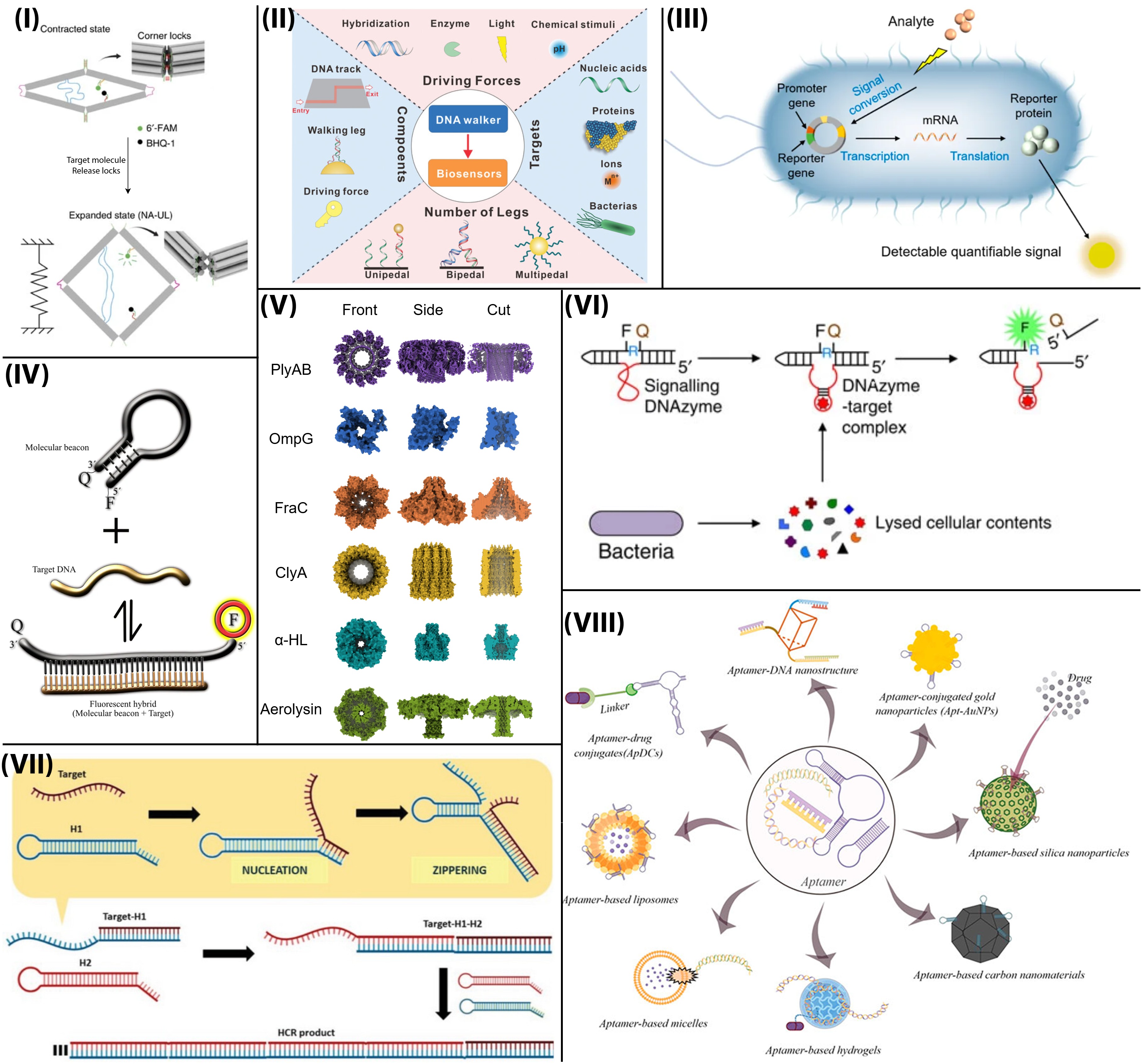}
  \caption{Overview of a selection of biosensors. (I) Schematic of a DNA-origami-based biosensor in which a 6-FAM fluorophore and a BHQ-1 quencher are held in close proximity until a target molecule releases the lock. Reproduced from~\cite{ke2016origamisensor}, licensed under CC BY 4.0. (II) Overview of driving forces, components, targets, and numbers of legs of DNA walking machines. Reproduced from~\cite{song2022dnawalker}, licensed under CC BY 4.0. (III) Visualization of bacteria-based biosensing. Reproduced from~\cite{gui2017wholecell}, licensed under CC BY 4.0. (IV) Concept of a DNA-hairpin sensor containing a quencher Q and a fluorophore F. Through strand displacement, the target strand opens the hairpin and separates quencher and fluorophore. Reproduced from~\cite{sampieri2022hairpin}, licensed under CC BY 4.0. (V) Different nanopores visualized with ChimeraX using structures from the Protein Data Bank (PDB). Depictions are not to scale. The selection is based on~\cite{mayer2022nanopores}. (VI) Concept of a DNAzyme-based biosensor. When bacteria break apart, they release cellular contents that activate the signalling DNAzyme. The activated DNAzyme cleaves its substrate, separating the fluorophore from the quencher and generating a fluorescence signal. Reproduced from~\cite{kang2014detection}, licensed under CC BY 4.0. (VII) Concept of the hybridization chain reaction. A target strand binds to the toehold of hairpin H1, initiating strand displacement and opening its stem. The exposed sequence then binds to the toehold of hairpin H2, triggering another strand displacement. This process repeats, resulting in a growing DNA assembly. Reproduced from~\cite{wong2025hcr}, licensed under CC BY 4.0. (VIII) Examples of different aptamer applications. Reproduced from~\cite{gao2022aptamers}, licensed under CC BY 4.0.}
  \label{fig:biosensor}
\end{figure*}

Within the overall DNA-based nanonetwork concept, the biosensor forms the detection stage that translates a specific detection event into a measurable response.
For the proposed proof-of-concept system, this requires biosensors capable of specifically detecting DNA strands.
Since biosensor designs differ in their detection mechanisms and specificity, this subsection examines selected candidates for the proposed proof-of-concept implementation.
There are many reviews, such as \cite{sahin2020sensors, wang2024sensors, liu2025sensors, shen2021sensors, ye2018sensors}, giving an overview of different types of biosensors.
The following paragraphs first introduce the considered biosensor implementations and their relevant properties before assessing their suitability according to the defined metrics.

A possible design principle for biosensors is DNA origami.
The DNA origami itself is not a recognition agent, but it serves as the building block of the sensor.
The concept is shown in Figure~\ref{fig:biosensor} (I) and is implemented with any recognition agent that can be integrated into DNA origami.
The nanobox with a controllable lid introduced in~\cite{andersen2009boxlid} could be one form of implementation.
The main concern is some sort of openable DNA origami structure that propagates a signal change based on the structures state.
The target could be a DNA strand that opens the DNA origami structure via strand displacement, but other target agents are feasible as well.
The exemplary DNA origami sensor in Figure~\ref{fig:biosensor} (I) holds a fluorophore and quencher in place while in its contracted state.
If the target molecule binds at the target sites, the corner locks open up and fluorophore and quencher are seperated.
DNA origami as a biosensor is considered as an highly adaptable implementation for signal amplification~\cite{loretan2020origamisensor, wang2020origamisensors}.

Another type of biosensor are DNA walking nanomachines.
An overview of DNA walking nanomachines can be seen in Figure~\ref{fig:biosensor} (II).
DNA walking machines can be implemented for different targets and with different driving forces, components and numbers of legs~\cite{cheng2021flap, song2022dnawalker}.
DNA walking nanomachines move along walking tracks using so-called walking strands, driven by enzymatic activity, strand displacement, light, or chemical stimuli.~\cite{song2022dnawalker}.
DNA walking machines can function as biosensors, if the walking strands hold a recognition agent.
The target binds to the walking strands and releases the recognition agent via strand displacement.
When the walking strands detect enough targets, the system produces a measurable reaction.
DNA walking nanomachines are a capable and versatile implementation of electrochemical or optical biosensors and are adaptable for many different sensing disciplines~\cite{shen2024walkingdna}.
The main limitation for our type of nanonetwork is the requirement for a predefined walking track, which is difficult to integrate into a system that relies on diffusion through a liquid medium.
Further research in terms of compatibility needs to be done, before considering DNA walking nanomachines for our setup.

O'Connor et al. reviewed bacteria as living platforms for DNA biosensing~\cite{oconnor2025bacteria}.
More generally, genetically engineered bacteria can function as whole-cell biosensors by converting the recognition of an analyte into a measurable output~\cite{gui2017wholecell}.
As illustrated in Figure~\ref{fig:biosensor}~(III), recognition of the analyte activates a regulatory element controlling a reporter gene.
The reporter gene is transcribed into mRNA and subsequently translated into a reporter protein that generates a detectable and quantifiable signal.
For DNA sensing, bacterial systems can additionally exploit mechanisms for recognizing extracellular DNA, including natural DNA uptake and engineered genetic or CRISPR-based detection mechanisms as reviewed by O'Connor et al.~\cite{oconnor2025bacteria}.
In this way, bacteria can translate the presence of specific DNA sequences into an experimentally measurable signal.

Figure~\ref{fig:biosensor}~(IV) shows an example of a hairpin DNA biosensor.
The hairpin DNA strand is a sensory concept for precise ssDNA targeting~\cite{tyagi1996hairpins}.
Its design is based on strand displacement.
The hairpin strand assembles into a loop and a stem, that is designed to get displaced by a more fitting strand of DNA.
The stem and loop are designed, such that a target strand binds with loop and stem and therefore displaces the connection of the stem and opens up the hairpin.
The hairpin sensor reaction is usually very hard to measure.
Additional signal amplification might be necessary, to ensure detection.

Nanopores for DNA detection are based on a two chamber setup containing an electrolyte solution, with both chambers separated by an electrically isolating lipid membrane holding a nanopore~\cite{mayer2022nanopores}.
The type of nanopore determines the target molecule which translocates through the nanopore.
Multiple different types of nanopores can be seen in Figure~\ref{fig:biosensor}~(V).
Using a motor protein to slow down the translocation, DNA is slowly fed through the nanopore.
By measuring the electric signal in both chambers and analyzing the changes in the current the exact base sequence can be reconstructed.
This is because Guanine, Adenine, Cytosine and Thymine have different sizes and forms and therefore modify the position of the ions and block the current between the chambers differently.

The biosensor conceptualized in Figure~\ref{fig:biosensor}~(VI) is DNAzyme-based~\cite{kang2014detection}.
The ability of DNAzymes to cleave specific strands can be used for biosensing.
A possible DNAzyme-based sensor holds a fluorophore and a quencher at an inactive DNAzyme and a so-called substrate strand~\cite{liu2025sensors}.
The target molecule binds at a recognition site and activates the DNAzyme, which cleaves the substrate strand.
The quencher and fluorophore get separated by this action and a fluorescent signal can be measured.
Other more complex implementations for biosensors are feasible by employing the cleaving mechanism of DNAzymes~\cite{guan2026dnazyme}.

\textit{Hybridization Chain Reaction} (HCR) is related to DNA hairpin sensors in its core mechanism.
The DNA hairpin sensors detect one target DNA strand and by that separate one quencher and fluorophore.
HCR uses two types of hairpins typically called H1 and H2 that both have a loop, a stem, and a toehold~\cite{wong2025hcr}.
A target strand binds to the toehold of H1 and opens the stem via strand displacement.
The now open strand of H1 can bind to the toehold of H2 and opens its stem again via strand displacement.
The open strand can bind to the H1 toehold again and thereby starts a cascade of H1 and H2 strands binding together into a long double helix.
This process can be seen in Figure~\ref{fig:biosensor} (VII)~\cite{wong2025hcr}.
This long double-helical strand can be used for signal transduction in combination with signaling agents such as silver nanoclusters~\cite{wong2025hcr}.

Figure~\ref{fig:biosensor}~(VIII) shows aptamers and their compatibility to different nanostructures.
Aptamers are short, single-stranded nucleic acids that bind to selected targets such as proteins, peptides, small molecules or ions~\cite{yang2023aptamer101}.
Aptamers are identified by the library selection process called \textit{Systematic Evolution of Ligands by Exponential Enrichment} (SELEX).
As of that they are highly specific in sensing target markers, up to single differences in amino acid~\cite{yang2023aptamer101}.
Other than most DNA structures, they can not be adapted because of that.
If the structure of the aptamer is changed, its sensory capability is changed~\cite{fallah2024aptamer}.
Because of that, other nanocomponents need to be adapted to the specific aptamer.
Their function is similar to antibodies, which are widely used as a molecular recognition agent.
Aptamers hold the following crucial merits over antibodies.
Aptamers are obtained in vitro by the SELEX technique~\cite{ettayri2025recent_aptamers}.
Antibodies need to be produced in complex cell structures or in vivo in animal experiments~\cite{ning2020aptamers}.
Because of that aptamers are obtained in a much shorter time and with less cost of construction.
Yang et al.\cite{yang2023aptamer101} compare antibodies and aptamers in both selection speed and commercial cost.
In their estimation Aptamers are obtained in days to weeks in vitro and cost around \SI{50}{\$\per\m\g}, based on the complexity of the chemical modifications.
Antibodies on the other hand take weeks in vivo and the production costs around \SIrange{2000}{5000}{\$\per\m\g}.
Also, aptamers can be more easily modified with additional functional groups through chemical adsorption or coupling~\cite{ning2020aptamers}.
This factor is of special importance for nanonetworks, because fluorophores, quenchers and nanomaterials have to be combined with the sensory system in order to have logical control.
As of the similarities of aptamers and antibodies, we only focus on aptamers in this work, because of their advantages in speed, cost and modification.

% green DCD8BC (5| 22-25)     \score{DCD8BC}{5}
% blue CDDBD8 (4 | 18-21)     \score{CDDBD8}{4}
% yellow ECDA88 (3 | 13-17)   \score{ECDA88}{3}
% orange F0CDB2 (2 | 9-12)    \score{F0CDB2}{2}
% red E8C0AD (1 | 5-8)        \score{E8C0AD}{1}

\begin{table}
  \centering
  \caption{Component comparison heatmap for biosensors in a minimal DNA-based nanonetwork PoC. The five parameters Amount of Subcomponents (AoS), Amount of Operations (AoO), Compatibility with DNA (CwD), Level of Stability (LoS) and Cost of Components (CoC) are scored 1--5.}
  \label{tab:component-biosensor-heatmap}
  \begin{tabular}{lrrrrrr}
    \toprule
    Component            & Score              & AoS               & AoO               & CwD               & LoS                & CoC                \\
    \midrule
    DNA hairpin          & \score{CDDBD8}{20} & \score{DCD8BC}{5} & \score{CDDBD8}{4} & \score{DCD8BC}{5} & \score{ECDA88}{3*} & \score{ECDA88}{3}  \\
    Aptamers             & \score{CDDBD8}{19} & \score{DCD8BC}{5} & \score{CDDBD8}{4} & \score{CDDBD8}{4} & \score{ECDA88}{3*} & \score{ECDA88}{3}  \\
    HCR                  & \score{CDDBD8}{19} & \score{DCD8BC}{5} & \score{CDDBD8}{4} & \score{DCD8BC}{5} & \score{ECDA88}{3*} & \score{F0CDB2}{2}  \\
    DNAzyme-based        & \score{ECDA88}{17} & \score{CDDBD8}{4} & \score{CDDBD8}{4} & \score{CDDBD8}{4} & \score{ECDA88}{3*} & \score{F0CDB2}{2}  \\
    Nanopores            & \score{ECDA88}{16} & \score{CDDBD8}{4} & \score{ECDA88}{3} & \score{DCD8BC}{5} & \score{ECDA88}{3}  & \score{E8C0AD}{1}  \\
    DNA origami-based    & \score{ECDA88}{15} & \score{E8C0AD}{1} & \score{CDDBD8}{4} & \score{DCD8BC}{5} & \score{CDDBD8}{4}  & \score{E8C0AD}{1}  \\
    DNA walking machines & \score{ECDA88}{15} & \score{E8C0AD}{1} & \score{CDDBD8}{4} & \score{DCD8BC}{5} & \score{CDDBD8}{4}  & \score{E8C0AD}{1}  \\
    bacteria-based       & \score{F0CDB2}{11} & \score{CDDBD8}{4} & \score{F0CDB2}{2} & \score{F0CDB2}{2} & \score{F0CDB2}{2*} & \score{E8C0AD}{1*} \\
    \bottomrule
    \footnotesize *low evidence
  \end{tabular}
\end{table}

We score the DNA hairpin sensors first.
The amount of subcomponents for DNA hairpin sensors is one, because quencher and fluorophore can be added easily to a DNA strand ordered from e.g. IDT~\cite{IDT2025}.
An AoS score of 5 follows.
The amount of operations can be seen as zero, because the complete sensor already exists from ordering the specific strand.
For the hairpin sensors and any DNA based sensor to function correctly in a wet-lab experiment, high base purity is important~\cite{zhang2016purification}.
For this purpose, purification methods such as \textit{high-performance liquid chromatography} (HPLC) or \textit{polyacrylamide gel electrophoresis} (PAGE) are necessary, which are done by DNA suppliers like IDT~\cite{IDT2025}.
But even the purified strands need to be initially annealed before using them for sensory tasks, to denature unwanted sequences and correctly form the hairpin loops~\cite{qin2023hairpin}.
A single annealing process right before the experiment therefore is necessary.
This results in an AoO score of 4.
DNA hairpin sensors are DNA native, which scores them a 5 for the compatibility with DNA~\cite{tyagi1996hairpins}.
We conclude the LoS from the stability analysis of ssDNA in Subsection~\ref{subsec:DNA_tiles}, which is 3.
The cost of DNA hairpin sensors needs to account for the necessity of clean and pure sequences.
For the cost analysis, we assume a 32-base-long hairpin strand with a base price of €18.56 from IDT~\cite{IDT2025}. Dual HPLC purification for \SI{100}{\nano\mole} adds €75, while fluorophore and quencher modifications are estimated at an additional €100. The resulting total cost of approximately €193.56 corresponds to a CoC score of 3.

The scoring for the amount of subcomponents for aptamers is 5, because the aptamer strand can be purchased and used as a single component.
The amount of operations is equivalent to DNA hairpin sensors with 4, taking one initial annealing step to denature unwanted sequences and forming the aptamers correctly.
Even though aptamers are DNA native, the compatibility with DNA is not that high for our use case.
Integrating aptamers into DNA structures is natively easy, but detecting specific DNA sequences is not given in their design.
The fact that aptamers need to be discovered via SELEX, makes them compatible to only a very specific amount of DNA strands.
We give aptamers a CwD score of 4 because of that.
The stability of aptamers is equal to ssDNA and therefore scored with 3.
Aptamers can be bought as ssDNA and need additional purification methods, just like hairpin sensors, to ensure a precise application of the sensors.
For the cost analysis, we consider the 41-nucleotide sgc8c aptamer, a shortened version of sgc8 targeting protein tyrosine kinase 7 (PTK7) on leukemia cells researched by~\cite{shangguan2008aptaprobe}.
The strand costs €23.78 from IDT, with an additional €75 for dual HPLC purification at \SI{100}{\nano\mole}.
Assuming another €100 for a fluorescence-based signaling mechanism, the estimated total cost is €198.78, resulting in a CoC score of 3.

The score for nanopores does not reflect the additional effort required for their experimental implementation.
Nanopores require an electrically isolated two-chamber setup with a small intersection spanned by an electrically insulating lipid membrane containing the nanopore~\cite{mayer2022nanopores}.
This additional experimental effort is not considered in the scoring, as the setup itself is included in neither the AoS nor the AoO complexity assessment.
The amount of subcomponents of nanopores is three with the pore, the lipid membrane and the electrolyte solution~\cite{mayer2022nanopores}.
The AoS score therefore is 4.
The amount of operations is 4 with building up the lipid membrane, including the nanopores, then spanning the membrane with the pore into the two-chamber setup and finally calibrating the setup with a baseline recording of the electrolyte solution~\cite{white2007nanopore}.
The AoO score for nanopores is 3.
Nanopores are not detecting a specific DNA strand.
The precise sequence of a DNA strand can be read from the change in current over time, which makes one nanopore able to detect any strand~\cite{wei2024nanopore}.
The limitation of needing the two-chamber setup does not matter for a proof-of-concept experiment, which results in a CwD score of 5.
The stability of nanopore biosensors mainly concerns the lipid membrane, which historically is unstable but can be optimized to be stable for many hours and relatively high voltages~\cite{tiefenauer2008lipidbilayer}.
Using a more stable lipid membrane results in a LoS score of 3.
The cost of nanopores from Oxford Nanopore Technologies starts at €740 for the flow cell kits~\cite{NanoporeTech2026}.
Further research into the cost of the special equipment needed for this type of biosensor is not necessary, because that kit alone results in a CoC score of 1.

DNAzyme-based biosensors usually consist of three subcomponents~\cite
{khan2021dnazyme}:
The DNAzyme, the substrate strand holding quencher and fluorophore and an aptamer or other targeting agent, that immobilizes the DNAzyme until the target molecule is bound.
This results in an AoS score of 4.
The amount of operations is equal to the other biosensors based on DNA before, with a single initial annealing process.
The AoO score therefore is 4.
The compatibility with DNA is based on the targeting agent, which often is an aptamer~\cite{khan2021dnazyme}.
The CwD score can be assumed as equal to the CwD score of aptamers with 4.
DNAzyme stability is equal to ssDNA and scored with 3.
The costs of DNAzyme-based biosensors can be generalized as follows.
Double HPLC costs €75 for each of the three tubes of strands from IDT, using a twelve-base-long substrate strand, a 40-base-long DNAzyme strand as well as a 41-base-long aptamer additional costs are €6,96, €23,20 and €23,78  for \SI{100}{\nano\mole} tubes~\cite{IDT2025}.
Including an estimated €100 for the fluorophore and quencher modifications of the substrate strand, the total cost is approximately €378.94, which corresponds to a CoC score of 2.

The amount of subcomponents for bacteria-based biosensors is based on the cultivation process.
Whole-cell bacterial biosensors commonly use engineered bacterial reporter strains, in which a sensing or stress-responsive element is coupled to a reporter gene~\cite{revilla2020wholecell,yin2022wholecell}.
For a plasmid-based implementation, this typically requires a cultivation medium, an initial bacterial culture, a plasmid-based reporter construct and a suitable selection antibiotic.
The resulting AoS score is 4.
The amount of operations is approximately six, although the exact workflow depends on the selected host organism and biosensor design~\cite{jing2026bacteria}.
We score the AoO with 2 because of that.
The compatibility with DNA is based on a single point of compatibility, because the bacteria needs to be cultivated with very specific receptor and analyte characteristics to work as a sensor for a specific strand of DNA~\cite{oconnor2025bacteria}.
The CwD score therefore is 2.
Bacteria-based biosensors are sensitive when cultivated in a wet-lab environment~\cite{gui2017wholecell}.
There are more stable bacteria cultures than others, but even those need specific temperature, pH values and environmental constraints to cultivate correctly~\cite{riangrungroj2019bacteria}.
We score the LoS as 2 because of that.
To estimate the costs of such a setup we look at one example from all four components needed, without necessarily picking compatible ones.
The medium could be EC-Bouillon for approximately €270 from Sigma Aldrich~\cite{sigmaaldrich2026}, the plasmid-based reporter structure could be pCRISPathBrick from addgene~\cite{addgene2026} for approximately €75, a common antibiotic such as Penicillin-Streptomycin from Sigma Aldrich~\cite{sigmaaldrich2026} for about €18 serves as a representative and a possible active bacteria culture could be E. coli from DSMZ~\cite{DSMZ2026} for €280.
Together this approximated bacteria kit costs €643 which results in a CoC score of 1.

DNA walking machines consist of many subcomponents, because the walking track is often built with DNA-origami.
Based on the DNA origami walking pad alone, the AoS score is 1.
For the amount of operations the DNA origami scores at 4 again, so does the whole setup.
The compatibility with DNA is native and therefore scored with 5.
DNA walking machines are stable duplex structures comparable to DX-tiles and the DNA origami track is highly stable as well, which results in a LoS score of 4~\cite{song2022dnawalker}.
The pricing of a relatively long DNA origami-based walking track results in prices well above €500, which gains a CoC score of 1 for DNA walking machines.

The scoring of the HCR is analogous to that of hairpin sensors, but uses two hairpins instead of one.
The AoS for HCR is scored with 5 because of the two hairpin strands holding quencher and fluorophore.
The AoO is an initial annealing process to denature unwanted sequences, which results in an AoO score of 4.
HCR is natively compatible with DNA and gets a CwD score of 5 because of that.
The level of stability is equal to hairpin sensors with a LoS score of 3.
The costs can be concluded from the costs of hairpin sensors as well.
Two hairpins for €193,56 each result in €387,12 and therefore in an CoC score of 2.

The scoring of DNA origami based sensors can be adapted one on one from the scoring of DNA origami cages, because only quenchers and fluorophores need to be added to a DNA origami cage to function as a biosensor.
This results in an AoS score of 1, an AoO score of 4, a CwD score of 5, a LoS score of 4 and a CoC score of 1.

The highest scoring DNA biosensor according to Table~\ref{tab:component-biosensor-heatmap} is a basic hairpin DNA sensor.
The only concern is the stability of the hairpin strands, but all the relatively high scoring biosensors have an approximately equal LoS.
Aptamers or a double hairpin biosensor setup for HCR is feasible as well, but the low cost and low complexity of basic hairpin DNA sensors gain an advantage for an in vitro proof-of-concept experiment.

The analysis of DNA-tiles, nanocarriers, and biosensors provides a set of suitable components for a minimal proof-of-concept experiment.
Based on these results, the following section combines the selected component types into a concrete experimental blueprint and defines how they interact within the proposed DNA-based nanonetwork.

%%%%%%%%%%%%%%%%%%%%%%%%%%%%%%%%%%%%%%%%%%%%%%%%%%%%%%%%%%%%%%%%%%%%%%%%%%%%%%%%%%%%%%%%%%%%%%%
%%%%%%%%%%%%%%%%%%%%%%%%%%%%%%%%%%%%%%%%%%%%%%%%%%%%%%%%%%%%%%%%%%%%%%%%%%%%%%%%%%%%%%%%%%%%%%%
%%% CONSTRUCTION %%%%%%%%%%%%%%%%%%%%%%%%%%%%%%%%%%%%%%%%%%%%%%%%%%%%%%%%%%%%%%%%%%%%%%%%%%%%%%
%%%%%%%%%%%%%%%%%%%%%%%%%%%%%%%%%%%%%%%%%%%%%%%%%%%%%%%%%%%%%%%%%%%%%%%%%%%%%%%%%%%%%%%%%%%%%%%
%%%%%%%%%%%%%%%%%%%%%%%%%%%%%%%%%%%%%%%%%%%%%%%%%%%%%%%%%%%%%%%%%%%%%%%%%%%%%%%%%%%%%%%%%%%%%%%

\section{The Blueprint for a Proof of Concept}
\label{sec:blueprint}

\begin{figure}
  \centering
  \includegraphics[width=\linewidth]{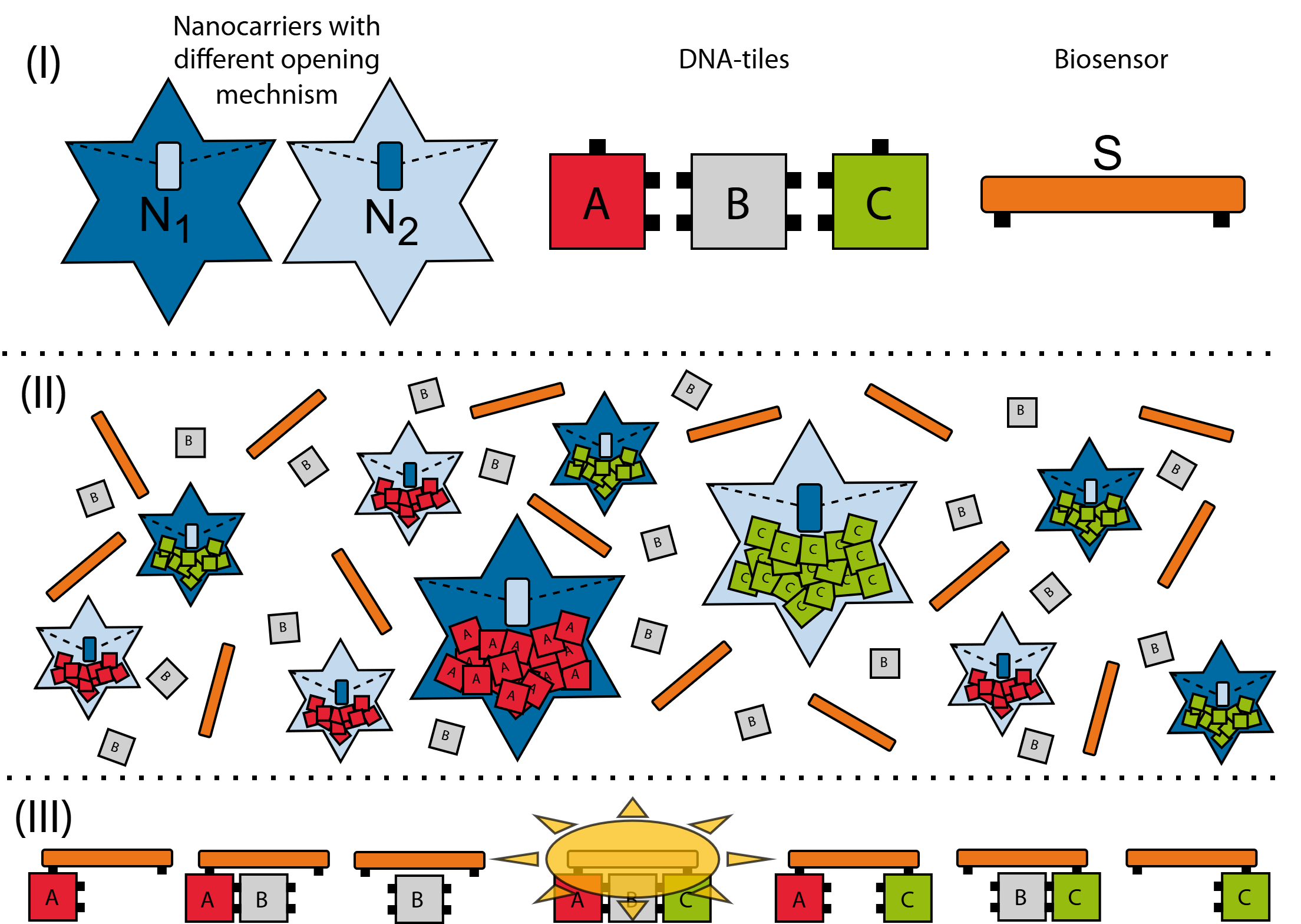}
  \caption{The minimal functional model defined for this paper. (I) The minimum component set. (II) The state of the medium for the model. (III) The intended reaction of the DNA-tiles with the biosensor. We assume, that the effect is fluorescence, but other measurable reactions are feasible.}
  \label{fig:model}
\end{figure}
Based on the concept of DNA-based nanonetworks introduced in Section~\ref{sec:prelim} and the component candidates analyzed in Section~\ref{sec:sota}, this section develops a concrete experimental blueprint for a proof-of-concept implementation.
The aim of this section is to integrate the selected DNA-tiles, nanocarriers, and biosensors into a coherent system architecture.
For that, we propose a minimal functional model of DNA-based nanonetworks, shown in Figure~\ref{fig:model}.
Figure~\ref{fig:model}~(I) shows the minimum set of components needed for a basic DNA-based nanonetwork.
DNA-tiles A, B and C together represent the message molecule that self-assembles in the nanonetwork.
We design A, B and C and their glues to bind with each other, while one glue of A and C is designed to bind with the biosensor.
The nanocarrier N$_1$ can be opened by biomarker M$_1$ and nanocarrier N$_2$ opens by biomarker M$_2$.
We design N$_1$ to hold DNA-tile A and N$_2$ to hold DNA-tile C as their cargo.
The biosensor S is able to detect the self-assembled message molecule from DNA-tiles A,B and C and delivers a measurable signal in case of a detection event.
Figure~\ref{fig:model}~(II) shows the state of the medium before adding the biomarkers M$_1$ and M$_2$.
Nanocarriers N$_1$ and N$_2$ hold their respective DNA-tiles A and C, DNA-tile B and the biosensor are freely moving in the medium.
Figure~\ref{fig:model}~(III) depicts the different self-assemblies possible in the experiment.
The expected fluorescent signal is only present, if the whole message molecule self-assembled beforehand.
If only one or two tiles are present in the medium, no signal should be detectable.

\begin{table*}
  \centering
  \caption{Oligos proposed for the blueprint with quencher 3IABkFQ and fluorophore 6-FAM.}
  \begin{tabular}{ll}
    \hline
    ID    & Sequence                                                                                              \\
    \hline
    A     & \texttt{5\textquotesingle-CTGGTGAATGGAAGAAGTCGCATCTTCCAT-3\textquotesingle}                           \\
    B     & \texttt{5\textquotesingle-ATCAGCGTTGAGACCTCTTCCATTCACCAG-3\textquotesingle}                           \\
    C     & \texttt{5\textquotesingle-GGTCTCAACGCTGATTGCGACTATCAGCGT-3\textquotesingle}                           \\
    G$_1$ & \texttt{5\textquotesingle-AATGCTAGCTAATCAATCGGG-3\textquotesingle}                                    \\
    K$_1$ & \texttt{5\textquotesingle-CCCGATTGATTAGCTAGCATT-3\textquotesingle}                                    \\
    G$_2$ & \texttt{5\textquotesingle-ACGTATTCGACTGACAGAGTA-3\textquotesingle}                                    \\
    K$_2$ & \texttt{5\textquotesingle-TACTCTGTCAGTCGAATACGT-3\textquotesingle}                                    \\
    S     & \texttt{5\textquotesingle-/6-FAM/ACGCTGATAGTCGCAGTATGGAAGATGCGACTATCAGCGT/3IABkFQ/-3\textquotesingle} \\
    \hline
  \end{tabular}
  \label{tab:oligos}
\end{table*}

Based on the highest-scoring nanocomponents from Section~\ref{sec:sota}, we design a minimal functional model.
We select SSTs as the DNA-tile implementation, for their high score of 22 out of possible 25 points last section.
The only concern is their stability, but in an in vitro experiment a low temperature constraint and an unpolluted medium help with that issue.
We select MSNs for the nanocarrier implementation, as to their high score in the last section.
The main concern of MSNs regarding their compatibility with DNA lies in their leakage problem and their highly discussed biocompatibility.
For the first problem we offer a tested solution, the second problem might concern in vivo experiment, but not necessarily the proof-of-concept experiment conducted in vitro.
We pick DNA hairpin sensors with quencher and fluorophore for the biosensor.

The hairpin sensor and SSTs are designed with typical DNA constraints in mind, but are not tested in their proposed form~\cite{hendling2019dna}.
For the nanocarrier implementation in this minimal proof-of-concept experiment, we use the MSN described by Climent et al.~\cite{climent2010msn}.
The MSNs are mesoporous MCM-41 nanoparticles with oligonucleotide guards to block the pores called G$_1$ and G$_2$ and their respective complementary key strands called K$_1$ and K$_2$ in our blueprint.
The strand pair G$_1$ and K$_1$ are equal to the pair presented in~\cite{climent2010msn}, G$_2$ and K$_2$ are designed to be orthogonal to G$_1$ and K$_1$ with the same sequence length.
In our PoC, the complementary key strands K$_1$ and K$_2$ serve as model triggers representing biomarkers M$_1$ and M$_2$, respectively.

\begin{figure}
  \centering
  \includegraphics[width=\linewidth]{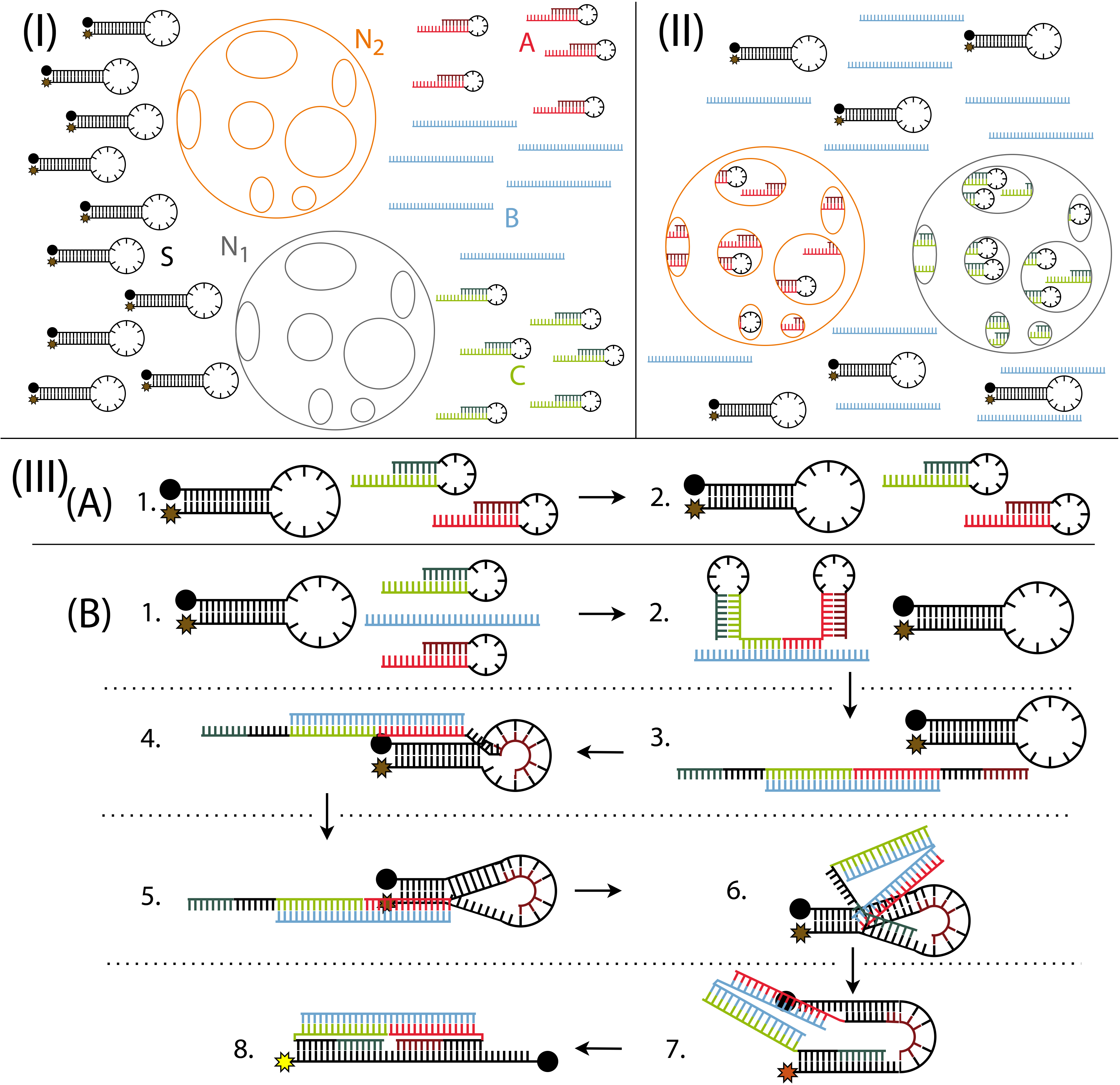}
  \caption{The proposed implementation of the model from Figure~\ref{fig:model} with highly scored components from Section~\ref{sec:sota}. (I) The proposed components with hairpin sensors S, MSNs N$_1$, N$_2$ and SSTs A, B, C. MSNs N$_1$ and N$_2$ need guard strands that block the pores, so that they do not leak their cargo. To reduce visual clutter, the guard strands are not depicted. (II) The state of the medium with the proposed components. (III) The proposed functionality of the SSTs and biosensor detection. (A) Without DNA-tile B the SSTs can not displace the hairpin sensor stem. (B) DNA-tile B displaces the stems of A and C. With that A and C can displace the stem of biosensor S.}
  \label{fig:implementation}
\end{figure}

Figure~\ref{fig:implementation} shows the proposed implementation of the model, using the same partitioning and labels as Figure~\ref{fig:model}.
Figure~\ref{fig:implementation}~(I) shows the components of the minimal set: the DNA hairpin sensors in black with quencher and fluorophore labeled S, the MSNs in orange and gray labeled N$_1$ and N$_2$, and the three SSTs in red, blue, and green labeled A, B, and C.
We design DNA-tiles A and C as hairpin strands to prevent strand displacement of the hairpin without tile B present.
The stem regions of both DNA-tiles A and C are the toeholds to displace the biosensor S.
The biosensors activate only if the stems are displaced beforehand.
We propose a ten base hairpin loop on the biosensor, with the toehold on the DNA-tile A with at least seven and at most ten bases to bind at the loop.
In the example of Figure~\ref{fig:implementation}~(I) the toehold of DNA-tile A has eight bases.
Figure~\ref{fig:implementation}~(II) shows the initial state of the implementation with both MSNs containing the SSTs A and C respectively.
The tile B and the hairpin sensor S are present in the medium as well.
In Figure~\ref{fig:implementation}~(III)~(A), we see the interaction of tiles A and C without tile B with the hairpin sensor.
Figure~\ref{fig:implementation}~(III)~(B) shows the interaction of all three tiles with the hairpin sensor.
We chose this mechanism and the structure of DNA-tiles and hairpins such that tile B is necessary for the activation of the hairpin sensor while tiles A and C displace the hairpin strands.
We design DNA-tiles A and C as hairpin strands with toeholds for DNA-tile B.
Tile B binds to both toeholds and displaces the stems of both DNA-tiles.
Figure~\ref{fig:implementation}~(III)~(B) 2 shows that process.
The rest of DNA-tile B then displaces the stems of DNA-tiles A and C in 3, resulting in a helix with two toeholds for the biosensor.
Figure~\ref{fig:implementation}~(III)~(B) 4 to 8 show the displacement process of both DNA-tiles A and C with biosensor S to seperate fluorophore and quencher.

The main concern of this implementation is the strand displacement of the hairpin sensor.
Without a functional strand displacement design a proof-of-concept experiment will fail.
The proposed design needs to be analyzed and evaluated in further research before employing the implementation.
We propose our strand design for all described components in Table~\ref{tab:oligos}.
Further, we design the sequences with consideration of guanine-cytosine balance, strand-displacement functionality and the intended hybridization interactions between DNA-tiles, guard and key strands, and biosensors.
However, the presented sequences represent an initial design proposal and require further in-silico validation before experimental implementation.
In particular, the sequences should be analyzed using tools such as NUPACK to predict secondary structures, hybridization equilibria, strand-displacement pathways, and potential off-target interactions~\cite{zadeh2011nupack}.
Such analyses are necessary to verify that the intended structures are thermodynamically favored and that undesired self-folding, cross-hybridization, or kinetic traps are minimized.
Therefore, the proposed sequences should be regarded as a conceptual design that requires comprehensive in silico validation and subsequent experimental optimization before wet-lab application.

\begin{figure}
  \centering
  \includegraphics[width=0.95\linewidth]{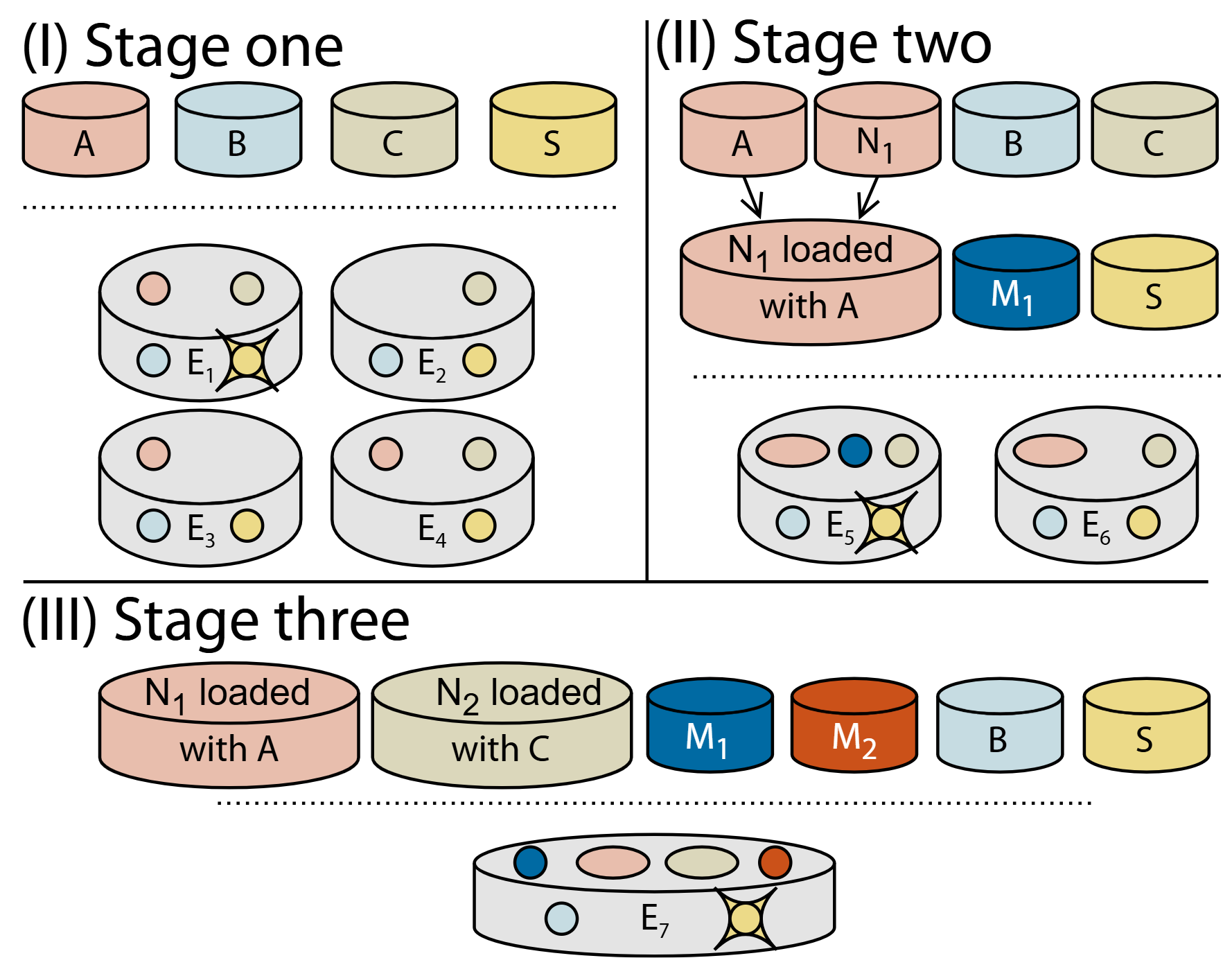}
  \caption{The experimental proof-of-concept blueprint. (I) Stage one: Proving the basic message assembly and detection mechanism. (II) Stage Two: Testing the functionality of the nanocarriers.  (III) Stage three: Combination of stage one and two with all components to test the whole DNA-based nanonetwork concept. We expect detection events for E$_1$, E$_5$ and E$_7$, while no detection events are expected for E$_2$, E$_3$, E$_4$ and E$_6$.}
  \label{fig:blueprint}
\end{figure}

We split the experimental proof-of-concept blueprint into three stages.
Stage one establishes the minimal PoC of DNA-tile-based self-assembly.
This is presented by the following logical formula:
\begin{align*}
  \mathrm{A} \land \mathrm{B} \land \mathrm{C}  \leftrightarrow \mathrm{S}
\end{align*}
For this, $A, B$ and $C$ is true, if the corresponding DNA-tiles are present in the experiment.
The required tests can be described with the following four logical statements:
\begin{align*}
  \mathrm{A} \land \mathrm{B} \land \mathrm{C}      & \rightarrow \mathrm{S}      \\
  \neg \mathrm{A} \land \mathrm{B} \land \mathrm{C} & \rightarrow \neg \mathrm{S} \\
  \mathrm{A} \land \neg \mathrm{B} \land \mathrm{C} & \rightarrow \neg \mathrm{S} \\
  \mathrm{A} \land \mathrm{B} \land \neg \mathrm{C} & \rightarrow \neg \mathrm{S}
\end{align*}

These four statements test the minimal functional requirement of our proof of concept: the complete three-tile assembly must trigger the biosensor, while removing any single tile must prevent detection.
Configurations with two or three missing tiles are not considered in this minimal experiment.
Additionally, this simple model does not consider the environmental conditions of the experiment, which are assumed to be constant and suitable for this blueprint.
Figure~\ref{fig:blueprint} (I) visualizes the four resulting experiments E$_1$, E$_2$, E$_3$ and E$_4$.
If all four statements can be proven experimentally, we have a minimal PoC of message molecules in DNA-based nanonetworks.
If stage one is successful, the second stage proves the functionality of the nanocarriers.
For that one nanocarrier needs to be loaded with one type of tile, for example N$_1$ loaded with A.
The chosen nanocarrier can be opened by adding biomarker M$_1$, which is tested with experiments E$_5$ and E$_6$ in Figure~\ref{fig:blueprint} (II).
With stage two, we prove the controlled release of DNA-tiles based on a specific biomarker.
If the nanocarrier works and the message molecule self-assembles correctly, one final experiment E$_7$ needs to be conducted in stage three, shown in Figure~\ref{fig:blueprint} (III).
For that, two different nanocarriers N$_1$ and N$_2$ need to be employed and loaded with tiles A and C respectively.
Nanocarrier N$_1$ releases tile A, if the biomarker M$_1$ is present and nanocarrier N$_2$ releases tile C, if the biomarker M$_2$ is present.
The loaded nanocarriers, the biomarkers, DNA-tile B and biosensor S in experiment E$_7$ are the PoC for the complete DNA-based nanonetworks as shown in Figure~\ref{fig:nanonetwork}.
All seven experiments with their employed components and the expected results are shown in Table~\ref{tab:experiments}.

\begin{table}
  \centering
  \caption{Proposed experiments and their used components DNA-tile A, DNA-tile B, DNA-tile C, nanocarrier N$_1$, nanocarrier N$_2$, biomarker M$_1$, biomarker M$_2$ and biosensor S. The presence of a component is depicted with \present, the absence with \absent. The result column shows if the expected result is a positive detection event (Det.) or if it is no detection event (No Det.) }
  \begin{tabular}{llllllllll}
    \toprule
    Exp.  & A        & B        & C        & N$_1$    & N$_2$    & M$_1$    & M$_2$    & S        & Result  \\
    \midrule
    E$_1$ & \present & \present & \present & \absent  & \absent  & \absent  & \absent  & \present & Det.    \\
    E$_2$ & \absent  & \present & \present & \absent  & \absent  & \absent  & \absent  & \present & No Det. \\
    E$_3$ & \present & \absent  & \present & \absent  & \absent  & \absent  & \absent  & \present & No Det. \\
    E$_4$ & \present & \present & \absent  & \absent  & \absent  & \absent  & \absent  & \present & No Det. \\
    E$_5$ & \present & \present & \present & \present & \absent  & \present & \absent  & \present & Det.    \\
    E$_6$ & \present & \present & \present & \present & \absent  & \absent  & \absent  & \present & No Det. \\
    E$_7$ & \present & \present & \present & \present & \present & \present & \present & \present & Det.    \\
    \bottomrule
  \end{tabular}
  \label{tab:experiments}
\end{table}
To realize the blueprint shown in Figure~\ref{fig:implementation}, we need eight batches for the initial annealing process:
\begin{enumerate}
  \item The SST A
  \item The SST B
  \item The SST C
  \item The biosensor S
  \item The guard strand used for the MSNs G$_1$
  \item The guard strand used for the MSNs G$_2$
  \item The key strand used for the MSNs K$_1$
  \item The key strand used for the MSNs K$_2$
\end{enumerate}
All these batches need to be heated up and cooled down separately to denaturize unwanted sequences and bind in the intended hairpin loops.
Parts of the batches of DNA-tiles A and C are mixed together with their respective MSN, to load the nanocarriers.
Loading MSNs can be achieved by simply mixing the cargo and MSNs over several minutes.
MSNs are leaking their cargo without additional guards, that block the pores.
Therefore, the loaded nanocarriers N$_1$ and N$_2$ need to be mixed with the guard strands G$_1$ and G$_2$ respectively, to block the pores.
After that step, we have eight batches with DNA-tiles A, B, C, biosensor S, nanocarriers N$_1$, N$_2$, key strands K$_1$ and K$_2$.
Both nanocarriers now hold their respective DNA-tiles A and C and the pores of the nanocarriers are blocked by the guard strands G$_1$ and G$_2$.
The batches can now be split into seven experiments, as depicted in Table~\ref{tab:experiments} and Figure~\ref{fig:blueprint}.

To obtain a detectable fluorescence signal from the biosensors, we choose a relatively high working concentration of \SI{100}{\nano\mole\per\liter}, based on molecular beacon assay conditions reported by Machinek et al.~\cite{machinek2014strands}.
All soluble DNA components are assumed to be used at this concentration in a final assay volume of \SI{200}{\micro\litre}.
The required molar quantity $n$ can then be estimated from the concentration $c$ and volume $V$ according to
\begin{align}
  n = c \cdot V = \SI{100}{\nano\mole\per\liter} \cdot \SI{200}{\micro\litre} = \SI{20}{\pico\mole}
\end{align}
per oligonucleotide species and experiment.
To account for experimental variability and to improve the statistical robustness of the results, three technical replicates are assumed for each experiment.
The required amount of each oligonucleotide can therefore be estimated from its number of occurrences in Table~\ref{tab:experiments}.
Table~\ref{tab:oligo_prices} summarizes the resulting minimum required quantities based on their usage across the experiments, a suitable order scale from IDT, the expected yields after purification, and the corresponding estimated ordering costs based on the purification method and IDT's pricing~\cite{IDT2025}.

\begin{table}
  \centering
  \caption{Overview of all oligonucleotides proposed in Table~\ref{tab:oligos} with their occurrences in Table~\ref{tab:experiments} and the resulting minimal required amount of molecules, a fitting order scale and the yield based on the purification technique and a price based on the strand lengths and IDT's pricing~\cite{IDT2025}. Tiles A, B and C are assumed to use Dual-HPLC and biosensor S is assumed to be ordered as a so-called Prime Time Dual-Labeled Probe.}
  \begin{tabular}{llrrrrr}
    \toprule
    ID    & Pur. & Occ. & Amount               & Order Scale          & Yield                 & Price   \\
    \midrule
    A     & yes  & 6    & \SI{360}{\pico\mole} & \SI{250}{\nano\mole} & \SI{6.5}{\nano\mole}  & €150,50 \\
    B     & yes  & 6    & \SI{360}{\pico\mole} & \SI{250}{\nano\mole} & \SI{6.5}{\nano\mole}  & €150,50 \\
    C     & yes  & 6    & \SI{360}{\pico\mole} & \SI{250}{\nano\mole} & \SI{6.5}{\nano\mole}  & €150,50 \\
    G$_1$ & no   & 3    & \SI{180}{\pico\mole} & \SI{25}{\nano\mole}  & \SI{10}{\nano\mole}   & €7,14   \\
    K$_1$ & no   & 2    & \SI{120}{\pico\mole} & \SI{25}{\nano\mole}  & \SI{10}{\nano\mole}   & €7,14   \\
    G$_2$ & no   & 1    & \SI{60}{\pico\mole}  & \SI{25}{\nano\mole}  & \SI{10}{\nano\mole}   & €7,14   \\
    K$_2$ & no   & 1    & \SI{60}{\pico\mole}  & \SI{25}{\nano\mole}  & \SI{10}{\nano\mole}   & €7,14   \\
    S     & yes  & 7    & \SI{420}{\pico\mole} & \SI{250}{\nano\mole} & \SI{12.5}{\nano\mole} & €268,00 \\
    \bottomrule
  \end{tabular}
  \label{tab:oligo_prices}
\end{table}

These quantities represent theoretical minimum requirements and do not account for pipetting losses, annealing optimization, gel validation, unsuccessful preparations, or repeated measurements.
The smallest synthesis scale offered by IDT is \SI{25}{\nano\mole}.
Based on the strand-displacement functionality discussed in Section~\ref{sec:sota}, DNA-tiles A, B and C represent critical functional components and are therefore assumed to be synthesized using Dual-HPLC purification.
Similarly, biosensor S is assumed to be ordered using IDT's \textit{PrimeTime Dual-Labeled Probe} option to obtain the required fluorophore and quencher modifications~\cite{IDT2025}.
According to the supplier specifications, these purification procedures substantially reduce the final yield compared to the nominal synthesis scale.
For example, a \SI{100}{\nano\mole} Dual-HPLC synthesis provides a guaranteed yield of approximately \SI{2}{\nano\mole}.
To provide sufficient material for optimization, characterization, and repeated measurements, a synthesis scale of \SI{250}{\nano\mole} therefore is proposed for DNA-tiles A, B and C as well as biosensor S.

For the final cost estimation, we select commercially available MCM-41 silica nanoparticles with a nominal diameter of 100 nm as a representative MSN platform.
According to Sigma-Aldrich, the corresponding material costs approximately €282.00~\cite{sigmaaldrich2026}.
Combined with the oligonucleotide costs summarized in Table~\ref{tab:oligo_prices}, this results in estimated material costs of approximately €1.030,06 for the proposed proof-of-concept implementation.
This estimate only includes the primary DNA components and nanocarriers and does not account for additional laboratory consumables, buffers, characterization measurements, or equipment-related costs.

%%%%%%%%%%%%%%%%%%%%%%%%%%%%%%%%%%%%%%%%%%%%%%%%%%%%%%%%%%%%%%%%%%%%%%%%%%%%%%%%%%%%%%%%%%%%%%%
%%%%%%%%%%%%%%%%%%%%%%%%%%%%%%%%%%%%%%%%%%%%%%%%%%%%%%%%%%%%%%%%%%%%%%%%%%%%%%%%%%%%%%%%%%%%%%%
%%% DISCUSSION %%%%%%%%%%%%%%%%%%%%%%%%%%%%%%%%%%%%%%%%%%%%%%%%%%%%%%%%%%%%%%%%%%%%%%%%%%%%%%%%
%%%%%%%%%%%%%%%%%%%%%%%%%%%%%%%%%%%%%%%%%%%%%%%%%%%%%%%%%%%%%%%%%%%%%%%%%%%%%%%%%%%%%%%%%%%%%%%
%%%%%%%%%%%%%%%%%%%%%%%%%%%%%%%%%%%%%%%%%%%%%%%%%%%%%%%%%%%%%%%%%%%%%%%%%%%%%%%%%%%%%%%%%%%%%%%

\section{Discussion}
\label{sec:disc}
%todo im dokument nach ?? suchen
Based on the preceding analysis of DNA-tiles, nanocarriers, and biosensors, we derived a blueprint for a minimal proof-of-concept experiment of a DNA-based nanonetwork.
This section discusses the implications and limitations of this blueprint and evaluates how it contributes to translating theoretical nanonetwork models into experimentally realizable systems.
The goal of this paper is to bridge the gap between the theoretical models of DNA-based nanonetworks and their physical realization in a wet-lab environment.
Previous work by Lau et al.~\cite{lau2019computation} and Gomez et al.~\cite{gomez2024abnormalitydetection} has established the logical correctness of DNA-tile-based computation and anomaly detection in nanonetworks.
However, logical correctness alone does not imply physical feasibility.
The transition from a formally verified model to a functioning wet-lab experiment is complex and requires many steps.
DNA nanotechnology has a long history of designs that are thermodynamically sound in simulation but fail in practice due to unintended secondary structures, kinetic traps, or incompatibilities between components that were individually validated but never tested in combination~\cite{badelt2020peppercorn, zhan2023dnaorigami}.
The proposed PoC addresses this gap.
%Todo: PoC mit passendem Begriff ersetzen
It is the first blueprint designed to prove, in stepwise wet-lab experiments that build upon one another, that the combination of SST tiles, a nanocarrier with a biomarker-controlled release mechanism and a hairpin-based biosensor can function as an integrated alarm system with message transmission in a DNA-based nanonetwork.
Each individual component has been experimentally validated in isolation in the literature, as discussed in Section~\ref{sec:sota}.
The blueprint presented in Section~\ref{sec:blueprint} aims to demonstrate that the component-based model and the implementation work together as planned, even in the laboratory.
%Todo:Begriffe richtig?

First, we discuss the scoring criteria and their underlying design decisions, as they provide the basis for the component selections made in this work.
Before the wet-lab experiments can be conducted, several assumptions in the blueprint require validation beforehand.
Some of these can be examined in-silico before any material is ordered.
We discuss those open design requirements, followed by a thorough analysis of the three-stage experimental design, to ensure that the proposed experiments are valid and feasible.
Further, we review the function of each experiment and explore possible failure modes and modifications.
Finally, cost considerations and the limitations of this paper are addressed to provide a more complete understanding of the proposed blueprint and its implications for future research in DNA-based nanonetworks.

\subsection{Scope and Transferability of the Scoring Framework}

We optimized the scoring criteria defined in Section~\ref{sec:sota}, amount of subcomponents, amount of operations, compatibility with DNA, level of stability, and cost of components, for an in vitro PoC with minimal cost and complexity.
We consider this a deliberate design decision rather than a limitation.
The framework is also adaptable to other scenarios and use cases.
For example, if in vivo suitability were introduced as an additional criterion, DNA origami cages and liposomes would score significantly higher due to their established biocompatibility and controlled release behaviour in biological media.
Similarly, if scalability to more complex message molecules were prioritized, crisscross structures would rank more favourably due to their modular extensibility~\cite{minev2021crisscross}.
The framework thus provides a structured basis for component selection that can be adapted to different experimental objectives and does not claim to be universally fitting.

This adaptability is particularly relevant for a computer science-oriented research group approaching wet-lab work for the first time.
Biochemical literature describes reagents in terms of synthesis protocols, reaction conditions, and molecular properties that do not map naturally onto the abstractions used in nanonetwork models.
A hairpin biosensor is not primarily a logical gate, and an MSN nanocarrier is not inherently a conditional release mechanism.
Bridging these two domains requires a structured method for comparing the relevant properties of individual constituents, and the scoring framework provides this basis.
By reducing the properties of each candidate to five criteria relevant to the experimental objective, it enables a systematic comparison that remains accessible to non-specialists.

The criteria are chosen carefully: AoS and AoO together capture the experimental complexity that a non-specialist group must manage; CwD captures the interoperability between parts that is essential in a multi-component system; LoS ensures that the chosen reagents remain stable throughout the experiment; and CoC reflects the practical reality that experiments must be repeated until a reliable protocol is established, meaning costs can quickly exceed the set budget if expensive reagents are selected.

Nonetheless, the scoring has its limitations.
A component that scores well on all five criteria for an in vitro PoC may score poorly if the experimental context changes.
A shift towards in vivo conditions would require additional criteria such as biocompatibility and nuclease resistance that are not captured by the current scoring.
Similarly, a focus on more complex message molecules would require scalability criteria that would favor different network architectures.
It should also be noted that the scoring reflects the state of the literature at the time of writing, and the individual implementation of each part may introduce additional complexity not captured by the score alone, as discussed in the preceding sections.
Further, the CoC values should be interpreted as approximate procurement cost categories rather than exact market prices.
Only one reference supplier or commercially available product was considered for each cost estimate, and prices may differ between suppliers, countries, package sizes, and purchase dates.
This specificity forces an explicit formulation of the experiments objective before any component selection is made. But that's a crucial step in the design of any interdisciplinary experiment.

\subsection{Open Design Requirements}

Several assumptions in the proposed blueprint require experimental or computational validation before the PoC can be conducted.
%todo blueprint conducted wäre doof
The most critical open question concerns the hairpin sensor. It is unclear, whether tile A, once released from the nanocarrier, is able to open the stem of the hairpin via toehold-mediated strand displacement.
If the toehold region of tile A does not bind to the loop of the hairpin with sufficient affinity, tile C would have no accessible toehold, and no fluorescence signal would be produced.
The thermodynamic feasibility of this displacement step can be evaluated using NUPACK~\cite{zadeh2011nupack} or similar tools prior to wet lab work.
In particular, a toehold length of at least 7~\textit{nucleotides} (nt) is recommended, as shorter toeholds have been shown to result in significantly lower displacement efficiencies~\cite{xu2013strand}.

Furthermore, the nanocarriers in the example of Climent et al.~\cite{climent2010msn} were loaded with fluorescein as cargo, not with DNA oligonucleotides.
It is therefore unclear, whether SST tiles can be loaded into the pores of MCM-41 MSNs under the same conditions, especially with the duality of DNA strands as cargo and as blocker at the same time.
The pore diameter of MCM-41 is approximately 3--4~nm, which is comparable to the estimated hydrodynamic radius of a 30~nt SST tile.
Whether the tiles can enter and be retained in the pores without additional surface functionalization, such as amino groups or polyethylenimine coating, is an open question that requires experimental validation as a prerequisite to stage two of the experimental blueprint.
Finally, the orthogonality of the guard and key strand pairs G$_1$/K$_1$ and G$_2$/K$_2$ is assumed but not yet verified.
The cross-hybridisation energies of G$_1$+K$_2$ and G$_2$+K$_1$ can be checked using NUPACK~\cite{zadeh2011nupack} or similar tools to confirm that no unintended strand exchange occurs between the two nanocarrier systems.

By resolving these open questions in advance, potential sources of systematic failure can be identified before the wet-lab phase begins.
The stagewise experimental design described in the following section is structured with these prerequisites in mind.

\subsection{Three-Stage Experimental Design}

The three-stage setup of the wet-lab experiment design reflects a deliberate separation of concerns that follows directly from the scoring criteria established in Section~\ref{sec:sota}.
The core argument is that stage one and stage two can fail independently and for different reasons.
Failure points of stage~1 are if the tile assembly does not proceed correctly, if tile A fails to open the hairpin stem, or when the blocking strand logic is not physically implemented as intended.
Stage two can fail because the MSN pores cannot be loaded with SST tiles, when the guard strands do not hold without the key strand present, or if the strand displacement release mechanism does not reach the concentrations required for detection.
If stage 3 with experiment E$_7$ were conducted without first validating stage one and stage two separately, a negative result would provide no actionable information about which part of the system failed.
The segmented, step-by-step design enables cost-efficient work, as each potential source of error can be identified in isolation and then addressed independently.

An additional advantage of stage 1 is the predictability of the concentrations of the components used.
The working concentration for SST tile experiments is typically in the range of 100--200~nM per strand species~\cite{wei2012sst}, which corresponds to approximately 20~pmol per experiment in a standard 200~\si{\micro\liter} assay volume.
For the hairpin sensor to produce a reliable fluorescence signal, working concentrations of 100~nM are typically required for direct detection without amplification~\cite{machinek2014strands}.
In stage one, this concentration is directly controlled by the experimenter.
In stage two and stage three, it depends on the loading capacity of the MSN nanocarrier and the efficiency of the strand displacement release, neither of which is known a priori for SST tiles as cargo.
Validating the detection system at known concentrations in stage one therefore also provides the baseline against which the release efficiency of the nanocarrier in stage two can be quantified.

Among the individual experiments every variant gives a meaningful insight.
E$_1$ proves that the model is valid, and the components are actually able to open the hairpin sensor and give a functional signal that can be detected.
A false negative would suggest that there is a mistake in the binding processes between A, B, C and the sensor.
The positive result in E$_1$ however needs to be investigated further, to prove that all components assemble correctly and are in fact necessary for a correct alert.
If a false positive result occurs in any of the experiments E$_2$, E$_3$, or E$_4$, each experiment points to a different source of error and has different implications for the redesign, that are discussed in the following paragraph.
Any false positive or false negative result needs to be investigated in detail, as other factors, such as contamination or environmental conditions, can also lead to unexpected results.

%todo musst sagen ob dir das AND da zu sehr herforgehoben ist
A core logical requirement of the proposed system is an \textsc{AND} relationship between tiles A and C, proving that both biomarkers M$_1$ and M$_2$ were detected before a signal is generated.
If experiment E$_2$, in which tile A is absent, produces a positive signal, tile C alone together with tile B is sufficient to open the hairpin.
The presence of M$_2$ alone would then be enough to generate a signal regardless of whether M$_1$ was detected, which would invalidate the specificity of the system.
The symmetric case applies to experiment E$_4$. A positive signal without tile C would mean that M$_1$ alone suffices, and the system can no longer distinguish between the detection of one or both biomarkers.
In both cases, the \textsc{AND} relationship between A and C is broken.
%todo: hier bin ich mir am unsichersten
The solution in either case would be to reduce the toehold length or binding energy of the affected tile.

Experiment E$_3$ addresses a different and equally important question. Even if E$_2$ and E$_4$ are negative, a positive signal without tile B would indicate that tiles A and C can possibly open the hairpin independently of one another, without ever forming a joint assembly.
This would undermine a central premise of the proposed nanonetwork.
The message molecule is not intended to consist of independently acting components. Rather, it is meant to form a controlled structure, created through self-assembly, comprising multiple building blocks.
In a more complex system, a message would be encoded in the specific arrangement of several interconnected tiles, and it would then be necessary to ensure that only the fully assembled message triggers a signal.
If individual tiles can activate the sensor on their own, there is no physical evidence that an assembly took place at all.
E$_3$ is therefore not just a negative control for the role of tile B in the displacement mechanism, but the experiment that establishes whether the system actually produces a message molecule or only a collection of reactive strands.

In stage two, the nanocarrier loading and release functionality is tested in isolation from the tile assembly logic of stage one.
Experiment E$_5$ tests whether the nanocarrier can be loaded with tile A and whether M$_1$ triggers its release.
If E$_5$ yields a false negative, this implies that either the loading mechanism is defective, meaning tile A is not retained in the nanocarrier under experimental conditions, or that the strand displacement release mechanism fails to produce sufficient concentrations of free tile A for detection.
As discussed above, the pore diameter of MCM-41 and the hydrodynamic radius of a 30~nt SST tile are comparable, and the loading capacity for SST tiles as cargo has not been established in the literature.
A false negative in E$_5$ should therefore first be investigated by testing the loading step independently before attributing the failure to the release mechanism.

Experiment E$_6$ tests the logical complement: whether the guard strands of nanocarrier N$_1$ hold in the absence of biomarker M$_1$, thereby validating the Climent et al.\ mechanism~\cite{climent2010msn} in the context of our design.
If tile A is released without M$_1$ present, the system cannot guarantee that nanocarrier N$_1$ only opens when the key biomarker has been detected.
The marker specificity of N$_1$ would be broken.
A corresponding experiment without M$_2$ should be included for nanocarrier N$_2$ to ensure symmetric validation of both release mechanisms, since the orthogonality of G$_2$/K$_2$ with respect to G$_1$/K$_1$ is assumed but not yet experimentally verified.

The marker logic of the full system depends on both nanocarriers maintaining their specificity independently.
If N$_1$ releases tile A only in the presence of M$_1$, and N$_2$ releases tile C only in the presence of M$_2$, then the \textsc{AND} relationship between M$_1$ and M$_2$ established in stage one between tiles A and C is physically realized at the level of the nanocarriers.
Stage two is therefore not only a validation of the nanocarrier mechanism in isolation, but a prerequisite for the integrity of the full system in stage three.
Experiment E$_7$ then brings all the components together.
The resulting positive outcome, supported by the preceding true-positive and correctly negative experiments, conclusively demonstrates that the implementation is functioning as intended.
Taken together, the seven experiments prove the necessity and sufficiency of all five components for the signal S, which corresponds exactly to the logical formula $\mathrm{N}_1(\mathrm{A}) \land \mathrm{N}_2(\mathrm{C}) \land \mathrm{M}_1 \land \mathrm{M}_2 \land \mathrm{B} \leftrightarrow \mathrm{S}$ introduced in Section~\ref{sec:blueprint}.

\subsection{Cost Considerations}

The cost estimates in Section~\ref{sec:sota} are based on the minimum purchasable quantities of each component and assume that the total amount can be divided across all experiments.
In practice, each try requires a sufficient amount of materials for reliable detection, and the actual cost per experiment may be two to three times higher than the per-component estimates suggest.
Nevertheless, the price of stage one can be reliably estimated on the high side, thanks to the more controllable concentrations. Therefore, the cost of stage one alone should be well within a few hundred euros.
Stage two and stage three introduce additional cost uncertainty due to the unknown loading capacity of the MSN nanocarrier for SST tile cargo, which cannot be estimated without prior experimental characterization.

\subsection{Limitations}

While the PoC is a significant step towards a functioning DNA-based nanonetwork as a detection, computation and communication system, we make no claim about the functionality of the system in biological media.
There, nuclease activity, ion concentrations, and temperature constraints differ substantially from the controlled in vitro environment, and many further steps are needed to translate the PoC towards in-situ or in vivo conditions.
Furthermore, it does not address the scalability of the approach to more complex messages, i.e. assemblies involving more than three tiles or more than two nanocarriers.
And it does not demonstrate biocompatibility of the chosen components, in particular the MSN nanocarrier, in a cellular environment.
These remain open questions for future work that builds on the foundation established here.
%%hier muss man dann jetzt in der conclusion wieder gut die kurve kriegen warum trotzdem alles super ist ;)

\section{Conclusion and Future Work}
\label{sec:conclusion}

This work was motivated by the overarching goal of bringing theory in DNA-tile based nanonetworks into the wet-lab.
While numerous simulation approaches for DNA-tile-based nanonetworks exist, standardized methods for transferring these concepts into experimental systems remain limited.
Wet-lab experiments are therefore a key next step toward bringing nanonetworks into real-world applications.

To address this gap, we first established the theoretical foundations of DNA-tile-based nanonetworks in Section~\ref{sec:prelim}.
We then analyzed possible implementations of DNA-tiles, nanocarriers, and biosensors and evaluated their suitability for a minimal proof-of-concept experiment in Section~\ref{sec:sota}.
Based on these findings, we developed a concrete experimental blueprint in Section~\ref{sec:blueprint}.
This provides a structured and reproducible methodology for selecting suitable components and developing experimental blueprints based on theoretical nanonetwork architectures.
Lastly, we discussed our findings and results in Section~\ref{sec:disc}.
This section summarizes our contribution and achievements, followed by the limitations and potential research directions in future.

\subsection{Summary of Contributions}

This work delivers the three contributions outlined in the introduction.
First, we identified possible components for DNA-tile based nanonetworks, namely DNA-tiles, nanocarriers, and biosensors.
To this end, we examined selected implementations of these components based on published research and their practical availability.
Second, we developed a scoring matrix that weighs DNA compatibility, cost efficiency, as well as experimental feasibility for each component class.
This provides not only a strong foundation for our own research, but also a valuable resource for the broader community invested in DNA-based nanonetwork and molecular communication research.
Furthermore, the scoring matrix delivers alternative components, if the theoretically best component is not feasible in experiments.
Third, we applied the scoring matrix to select the most suitable components for a proof-of-concept experiment.
This resulted in a promising blueprint with alternative candidates for each component, if experimental complications occur.
In conclusion, we developed a reproducible pipeline to bring simulated nanonetwork architectures into wet-lab experiments, with a methodology to find the most suitable components for them.

\subsection{Future Work}
The immediate next step is to validate the proposed DNA sequences and their intended interactions.
This includes analyzing secondary structures and potential off-target interactions in silico, followed by experimental verification of the strand-displacement mechanism.
Furthermore, the loading, retention, and controlled release of DNA-tile cargo from MSNs must be characterized before the complete proof-of-concept system can be implemented.
For now, this work is focussing on minimal, often repeated wet-lab experiments.
Therefore, the network architecture, component analysis, and scoring matrix are focused on that goal.
Nevertheless, we are confident that if the focus of the experiments shifts towards more complex networks or refined experimental setups, this work can be adapted due to its modularity.
Also, its flexible structure means it can be readily extended as research setups mature.

An important open research question is whether the hairpin tile design generalizes into larger assemblies like complex message molecules.
Further research should compare this design to other tile architectures with regard to scalability and stability.
Likewise, biocompatibility, including the behavior of open strands in the human body, represents an important and promising avenue for future research.
Resolving component selection can directly pave the way for targeted biocompatibility studies of the selected components.

\section*{Acknowledgments}
This research was supported in part by the project IoBNT, funded by the German Federal Ministry of Research, Technology and Space (BMFTR) under grant number 16KIS1991, and by the project NaBoCom III, funded by the German Research Foundation (DFG) under grant number 419981515.

\section*{Declaration of Generative AI Use}
We used OpenAI's ChatGPT (GPT-5 and GPT-6) to improve the clarity, grammar, and readability of the manuscript. The authors reviewed the suggested changes and remain responsible for the scientific content.

\printcredits

\bibliographystyle{elsarticle-num}
%\bibliographystyle{cas-model2-names}

\bibliography{cas-refs}



\end{document}